\documentclass[twocolumn]{aastex7}
\usepackage{amsmath, balance}
\usepackage{colortbl, multirow}
\usepackage{booktabs}
\usepackage{natbib}
\usepackage{caption}
\usepackage{subcaption}
\usepackage{bm}
\usepackage{dcolumn}
\usepackage{wasysym}
\usepackage{enumitem}
\usepackage{tikz}

\usetikzlibrary{
    shapes.geometric, 
    arrows.meta,      
    positioning,      
    calc              
}

\graphicspath{{./fig/}}

\newcommand{\model}{\texttt{SwinYNet}}

\newcolumntype{d}[1]{D{.}{.}{#1}}

\begin{document}

\title{SwinYNet: A Transformer-based Multi-Task Model for Accurate and Efficient FRB Search}

\correspondingauthor{Huaxi Chen, Yi Feng, Di Li}

\author[0009-0000-4795-8767]{Yunchuan Chen}
\affiliation{Research Center for Computational Earth and Space Science, Zhejiang Lab, Hangzhou 311121, China}
\email{chenych@zhejianglab.org}

\author[0000-0002-5386-1627]{Shulei Ni}
\affiliation{Research Center for Computational Earth and Space Science, Zhejiang Lab, Hangzhou 311121, China}
\email{nisl@zhejianglab.org}

\author[0000-0002-6736-3557]{Chan Li}
\affiliation{School of Sciences, Hangzhou Dianzi University, Hangzhou 310018, China}
\email{chanli@hdu.edu.cn}

\author[0000-0001-9956-6298]{Jianhua Fang}
\affiliation{Research Center for Computational Earth and Space Science, Zhejiang Lab, Hangzhou 311121, China}
\email{fangjh@zhejianglab.org}

\author[0000-0002-7420-9988]{Dengke Zhou}
\affiliation{Research Center for Computational Earth and Space Science, Zhejiang Lab, Hangzhou 311121, China}
\email{zdk@zhejianglab.org}

\author[0009-0000-6108-2730]{Huaxi Chen}
\affiliation{Research Center for Computational Earth and Space Science, Zhejiang Lab, Hangzhou 311121, China}
\email[show]{chenhxtony@gmail.com}

\author[0000-0002-0475-7479]{Yi Feng}
\affiliation{Research Center for Computational Earth and Space Science, Zhejiang Lab, Hangzhou 311121, China}
\email[show]{yifeng@zhejianglab.org}

\author[0000-0002-3386-7159]{Pei Wang}
\affiliation{National Astronomical Observatories, Chinese Academy of Sciences, Beijing 100101, China}
\affiliation{Institute for Frontiers in Astronomy and Astrophysics, Beijing Normal University, Beijing 102206, China}
\affiliation{State Key Laboratory of Radio Astronomy and Technology, Beijing 100101, China}
\email{wangpei@nao.cas.cn}

\author[0009-0007-5730-6958]{Chenwu Jin}
\affiliation{Research Center for Computational Earth and Space Science, Zhejiang Lab, Hangzhou 311121, China}
\email{jincw@zhejianglab.org}

\author[0009-0000-4098-1016]{Han Wang}
\affiliation{Research Center for Computational Earth and Space Science, Zhejiang Lab, Hangzhou 311121, China}
\email{wanghan@zhejianglab.org}

\author[0009-0003-1822-1717]{Bijuan Huang}
\affiliation{Research Center for Computational Earth and Space Science, Zhejiang Lab, Hangzhou 311121, China}
\email{huangbj@zhejianglab.org}

\author[0009-0005-5541-9436]{Xuerong Guo}
\affiliation{Research Center for Computational Earth and Space Science, Zhejiang Lab, Hangzhou 311121, China}
\email{guoxr@zhejianglab.org}

\author[0000-0003-4811-2581]{Donghui Quan}
\affiliation{Research Center for Computational Earth and Space Science, Zhejiang Lab, Hangzhou 311121, China}
\email{donghui.quan@zhejianglab.org}

\author[0000-0003-3010-7661]{Di Li}
\affiliation{New Cornerstone Science Laboratory, Department of Astronomy, Tsinghua University, Beijing 100084, China}
\affiliation{National Astronomical Observatories, Chinese Academy of Sciences, Beijing 100101, China}
\affiliation{State Key Laboratory of Radio Astronomy and Technology, Beijing 100101, China}
\email[show]{dili@tsinghua.edu.cn}

\begin{abstract}
In this study, we present a transformer-based multi-task model for Fast Radio Burst (FRB) detection, signal segmentation, and parameter estimation directly from time–frequency data, without requiring computationally expensive de-dispersion preprocessing.
To overcome the scarcity of labeled observational data, we develop an FRB simulator and a rule-based automatic annotation pipeline, enabling training exclusively on simulated data.
Evaluations on the FAST-FREX dataset show that our model achieves an \textbf{F1 score of 97.8\%, recall of 95.7\%}, and \textbf{precision of 100\%}, outperforming both conventional tools (e.g., PRESTO, Heimdall) and recent AI-based baselines (e.g., RaSPDAM, DRAFTS) in both accuracy and inference speed. The model supports pixel-level signal segmentation and yields reliable estimates for dispersion measure (DM) and time of arrival (ToA).
Large-scale blind searches on CRAFTS data further demonstrate robustness, with an \textbf{average false positive rate of 0.28\%} and minimal human verification required. This search has already led to the identification of \textbf{two pulsar candidates}, both confirmed as known pulsars.
Processing benchmarks indicate that the model enables \textbf{real-time searches on a single consumer-grade GPU}, making petabyte-scale blind searches feasible.
The code is publicly available on GitHub, and the model can be easily integrated with existing tools to automate and streamline radio data analysis beyond FRB or pulsar searches.
\end{abstract}


\section{Introduction} \label{sec:intro}

The rapid growth of radio astronomy is creating an unprecedented demand for more capable and intelligent data-analysis tools.
Modern radio telescopes---such as the Five-hundred-meter Aperture Spherical radio Telescope (FAST)~\citep{Nan:2011um,Li:2012ub,li2018fast},

the Very Large Array~\citep{napier1983very,perley2011expanded,Kellermann2020},
MeerKAT~\citep{Camilo:2018hsu,10.1093/mnras/sty2785},
CHIME~\citep{Bandura:2014gwa,CHIMEFRB:2018mlh},
LOFAR~\citep{ROTTGERING2003405,5109710},
ASKAP~\citep{ASKAP:2007rlq,Johnston:2008hp},
and MWA~\citep{tingay2013murchison,5164979}---are generating enormous volumes of observational data to support a wide range of scientific objectives,
including FRB detection and characterization \citep{CHIMEFRB:2018mlh},
pulsar timing and gravitational-wave detection \citep{lommenPulsarTimingGravitational2017,NAP12982},
neutral hydrogen mapping \citep{Paul:2023yrr,CHIMEFRB:2018mlh},
and exploring magnetic fields in the Universe \citep{beckCosmicMagneticFields2011}.
New facilities currently under construction, including the QiTai radio Telescope (QTT)~\citep{Wang:2023occ} and the Square Kilometer Array (SKA)~\citep{Braun:2015zta}, are expected to further enhance these capabilities in the near future.

These endeavors are driving major astrophysical discoveries and continuously deepening our understanding of the Universe.
However, existing data analysis tools often fall short in terms of efficiency, accuracy, and usability, especially when faced with the scale and complexity of next-generation datasets. This growing gap underscores an urgent need for more advanced, intelligent algorithms capable of understanding radio astronomical data with greater speed and reliability, thereby accelerating scientific discovery and maximizing the scientific return from these powerful observatories.
In this context, deep learning–based approaches have emerged as a promising solution for FRB searches, enabling efficient and accurate analysis of large-scale radio astronomical datasets.

FRBs are astrophysically valuable yet notoriously difficult to detect, making their systematic study both compelling and challenging. They are a class of highly energetic, millisecond-duration radio transients of extragalactic origin, characterized by large  dispersion measure (DM)~\citep{lorimer2007bright, masui2015dense, petroff2019fast,feng2022frequency}.
Since their discovery, FRBs have attracted a lot of attention due to their potential in probing intergalactic matter, testing fundamental physics, and uncovering new astrophysical phenomena \citep{thornton2013population,chatterjee2017direct,cordes2019fast,Zhang:2022uzl}.
The very nature of FRBs---their short duration, unpredictable occurrence in the sky, and the challenges of distinguishing them from terrestrial radio frequency interference (RFI) and the dispersive effects of the interstellar medium---poses significant hurdles for their detection and comprehensive study. The difficulty in reliably identifying these fleeting signals has historically limited the size of FRB catalogs and consequently our understanding of their origins and the physical mechanisms that produce them~\citep{cordes2019fast,zhang2020physical}.

Although deep learning have shown considerable promise for FRB searches, current methods still face several fundamental limitations. First, there exists a trade-off between detection accuracy and computational efficiency: end-to-end models operating on raw data often achieve limited accuracy, whereas higher-performance approaches typically depend on conventional pipelines augmented with deep learning components, significantly increasing computational cost and limiting scalability.

Second, progress is constrained by the scarcity of large, well-labeled datasets of real FRBs. The rarity and transient nature of these signals make it difficult to assemble diverse training datasets from observations alone, particularly when fine-grained annotations, such as semantic segmentation, are required beyond simple classification.

Third, most deep learning models provide limited functionality and interpretability, as they typically output only coarse-grained predictions such as class labels or object bounding box. Implemented as black-box classifiers or object detectors, they offer minimal explanation for their predictions. As a result, extensive manual verification is required, undermining confidence in fully automated systems.

Finally, these models often suffer from limited availability and usability. Their tight coupling to specific pipelines or telescopes leads to poor transferability and reproducibility, further exacerbated by the reliance on intermediate pipeline products as training data, which are difficult to share.

This study presents a versatile deep learning framework designed to streamline the FRB search and analysis pipeline while addressing the aforementioned limitations. The main contributions of this work are summarized as follows:
\begin{itemize}[leftmargin=1em, topsep=4pt plus 0pt minus 2pt, itemsep=0pt, parsep=3pt minus 1pt]
    \item \textbf{Multi-task Learning with SOTA Performance:}
    We propose a model that simultaneously detects signals, performs segmentation, and estimates parameters directly from raw dynamic spectra. Our model achieves state-of-the-art accuracy (F1 score: \textbf{97.8\%}) on the FAST-FREX dataset while significantly accelerating the search pipeline by bypassing expensive preprocessing steps such as RFI mitigation and de-dispersion.

    \item \textbf{Automated Simulation-Based Training:}
    We introduce a rule-based automatic annotation method using FRB simulations to facilitate large-scale training. This approach eliminates the need for manual labeling and demonstrates robust generalization across the simulation-to-reality domain gap, making it readily adaptable to data from diverse radio telescopes.

    \item \textbf{Interpretability and Workflow Integration:} Our model outputs pixel-level masks that serve as robust diagnostic features, significantly enhancing the interpretability of detection results. By isolating signals within these masks, the framework enables precise DM and ToA estimation, which provides high-quality initializations for traditional tools such as \texttt{fitburst} and \texttt{prepfold}. This integration creates an automated pipeline that bridges deep learning with established radio astronomy workflows.

    \item \textbf{Petabyte-Scale Validation:}
    In a blind search of petabyte-scale data, the model identified two pulsar candidates with a remarkably low false-positive rate of \textbf{0.28\%}. This large-scale application proves the model's efficiency and reliability for high-volume astronomical surveys.
\end{itemize}

To ensure reproducibility and facilitate further research, we have made our model and code publicly available on GitHub\footnote{
\url{https://github.com/expnn/SwinYNet}}
and Zenodo \citep{chenSwinYNet2025}.


\section{Related Work}\label{sec:related_work}
\subsection{Traditional Methods for FRB Detection}

Traditional single-pulse search pipelines used as FRB search tools---such as  PRESTO~\citep{ransomNewSearchTechniques2001} and Heimdall~\citep{barsdellAcceleratingIncoherentDedispersion2012}---operate by de-dispersion data across a dense grid of DMs followed by matched filtering and thresholding.
These methods are conceptually intuitive and closely align with established empirical practices in data processing, and they continue to serve as the dominant approach in FRB searches.

However, these approaches incur substantial computational complexity, exhibit limited adaptability to faint bursts, remain highly susceptible to RFI contamination, and frequently produce large numbers of false positives as well as duplicated detections~\citep{keane2018future,petroff2019fast,rafiei2023mitigating}.
Consequently, these tools have become increasingly impractical for modern data volumes, making manual verification a critical bottleneck.

\subsection{Deep Learning Methods for FRB Detection}
Deep learning has increasingly been explored as a promising alternative to traditional FRB search pipelines, mitigating issues including computational inefficiency, high false-positive rates, duplicated detections, and missed events.

\paragraph{Detection \& Classification of Candidates}

An early direction in applying deep learning to FRB detection focused on the binary classification of candidate signals identified by traditional pipelines.
A pioneering effort in this domain was the work of \citet{connorApplyingDeepLearning2018}, which employed a tree-like deep neural network trained on a mixture of simulated FRBs, real false positives, and pulsar signals.
Subsequent work, such as FETCH \citep{agarwalFETCHDeeplearningBased2020} and intensityML \citep{yadavApplyingConvolutionalNeural2020}, utilized convolutional neural networks (CNNs) to classify candidates as either astrophysical or spurious.
These physics-informed methods explicitly utilize the $\nu^{-2}$ propagation delay \citep{lorimerBinaryMillisecondPulsars1998},
\begin{equation}
t(\nu) = k_{\mathrm{DM}} \cdot \mathrm{DM} \cdot (\nu^{-2} - \nu_0^{-2}) + t_0,
\label{eq:time-freq-dm}
\end{equation}
to perform de-dispersion trials in preprocessing stages,
where $k_{\mathrm{DM}} \approx 4.15 \times 10^6\,\text{MHz}^2\cdot\text{pc}^{-1}\cdot \text{cm}^3\cdot \text{s}$ is the dispersion constant, and $(t, \nu)$ denotes the frequency-dependent arrival time relative to a reference coordinate $(t_0, \nu_0)$.
This delay is a fundamental characteristic of FRBs caused by the propagation through ionized media.

These approaches reduce the manual burden of verifying large numbers of false positives and enhance pipeline efficiency. However, since they rely on candidates generated by conventional search algorithms, they inherit upstream limitations such as detection incompleteness, inefficiency, and dependency on predefined DM trials.

Despite these advantages, DRAFTS also has certain limitations.
It operates on time-DM images, which require de-dispersing the data across a dense grid of DM trials, resulting in substantial computational overhead.
Moreover, in wide-band observations, FRB pulses often exhibit complex frequency-dependent structures \citep{Gajjar:2018bth}, which are not effectively preserved in time-DM representations.
As a result, the model's ability to accurately identify burst signatures may be impaired.
These limitations indicate that, although DRAFTS represents an important step toward highly accurate FRB searches, its reliance on dedispersion-based preprocessing may limit scalability and robustness across diverse observational settings.

\paragraph{Direct Detection in Raw Data}
A more ambitious line of work applies deep learning models directly to raw time-frequency data to identify FRBs as \emph{structured patterns}, typically modeled by Eq.~\eqref{eq:time-freq-dm}. Notable examples include \citep{Zhang:2018jux} and \citep{liuSearchTechniqueBased2022}, where convolutional neural network (CNN)-based classifiers are trained to detect such features without relying on the expensive dedispersion trials.

While these classifier-based approaches are computationally efficient, they often exhibit poor recall for weak events and provide limited interpretability, which complicates human verification and confidence calibration in practical deployments. Moreover, they integrate poorly with established analysis workflows that require precise parameter estimates. In such cases, physical parameters must still be derived via conventional tools. However, the model provides little guidance for initializing these tools, which often require precise starting values to function effectively.

\citet{guoAcceleratingFRBSearch2024} proposed RaSPDAM, a FRB detection algorithm that formulates the detection task as a semantic segmentation problem.
The raw dynamic spectra are transformed into enhanced 2D images, which are processed by a U-Net segmentation model to identify candidate FRB regions.
Their method achieves high precision and improved inference speed compared to traditional pipelines such as PRESTO and Heimdall.

However, RaSPDAM relies on a set of handcrafted post-processing heuristics to filter segmentation outputs and determine whether a detection corresponds to an FRB pulse. These include geometric constraints based on region area, bounding box ratios, and signal extent in the time and frequency axes. Such rule-based filtering can be sensitive to parameter tuning and may not generalize well across data from different telescopes or under varying RFI environments. Furthermore, while the method provides accurate estimates of ToA, it does not directly yield DM, limiting its utility for complete parameter estimation and subsequent astrophysical analyses.

\subsection{FRB simulation to generate training data}
To overcome the data bottleneck in training deep learning models, researchers have turned to simulating FRB data for training deep learning models \citep{2025asi..confO..64K}.
Various methods exist for generating these simulations, ranging from simple parametric models \citep{Luo:2020wfx,fonsecaModelingMorphologyFast2024,Kuiper:2024wsi} to more complex, physically motivated approaches \citep{Kader:2024uqm,Kuiper:2024wsi}. These methods often involve injecting synthetic FRB signals into real radio data to create diverse and realistic datasets \citep{agarwalFETCHDeeplearningBased2020}.
Training on simulated datasets constitutes an implicit physics-informed strategy.
This approach represents physical laws in the form of training samples, enabling the model to internalize the domain physics via a data-driven learning paradigm \citep{liuSearchTechniqueBased2022}.
For training FRB search models, simpler parametric simulation methods are often sufficient, as they enable the efficient generation of large and diverse datasets, which is essential for developing robust deep learning models.

\section{Methods}\label{sec:methods}

\subsection{Overview and Design Principles}
The primary objective of our framework is to transform FRB search from a multi-stage process into a unified, high-throughput inference task. To achieve this, the system is built upon three core design principles:

\vspace{-1ex}
\paragraph{Supervised Classification for RFI Discrimination}
We adopt a binary classification framework rather than anomaly detection to explicitly suppress RFI. Since many terrestrial interferences (e.g., narrow-band spikes) are statistically ``anomalous'' yet morphologically distinct from FRBs, a pure anomaly detector would trigger excessive false alarms. By incorporating supervised negative samples, the model establishes a sharper discriminative boundary that effectively isolates genuine signals from complex RFI environments.

\vspace{-1ex}
\paragraph{Multi-Task Unified Inference}
A shared Transformer-based backbone performs detection, pixel-level segmentation, and DM estimation simultaneously. This integration minimizes computational overhead and leverages spatial segmentation features to regularize and refine the classification performance, ensuring both efficiency and robustness.

\vspace{-1ex}
\paragraph{Simulation-Driven Generalization}
To circumvent the absence of pixel-level labels in real-world data, the model is trained on physics-informed simulations. By capturing universal invariants, such as the $\nu^{-2}$ dispersion sweep according to Eq.~\eqref{eq:time-freq-dm}, the framework achieves robust telescope-independent generalization without requiring any manual annotation.

\subsection{Architecture}
\begin{figure}[!tbp]
    \centering
    \includegraphics[width=0.99\linewidth]{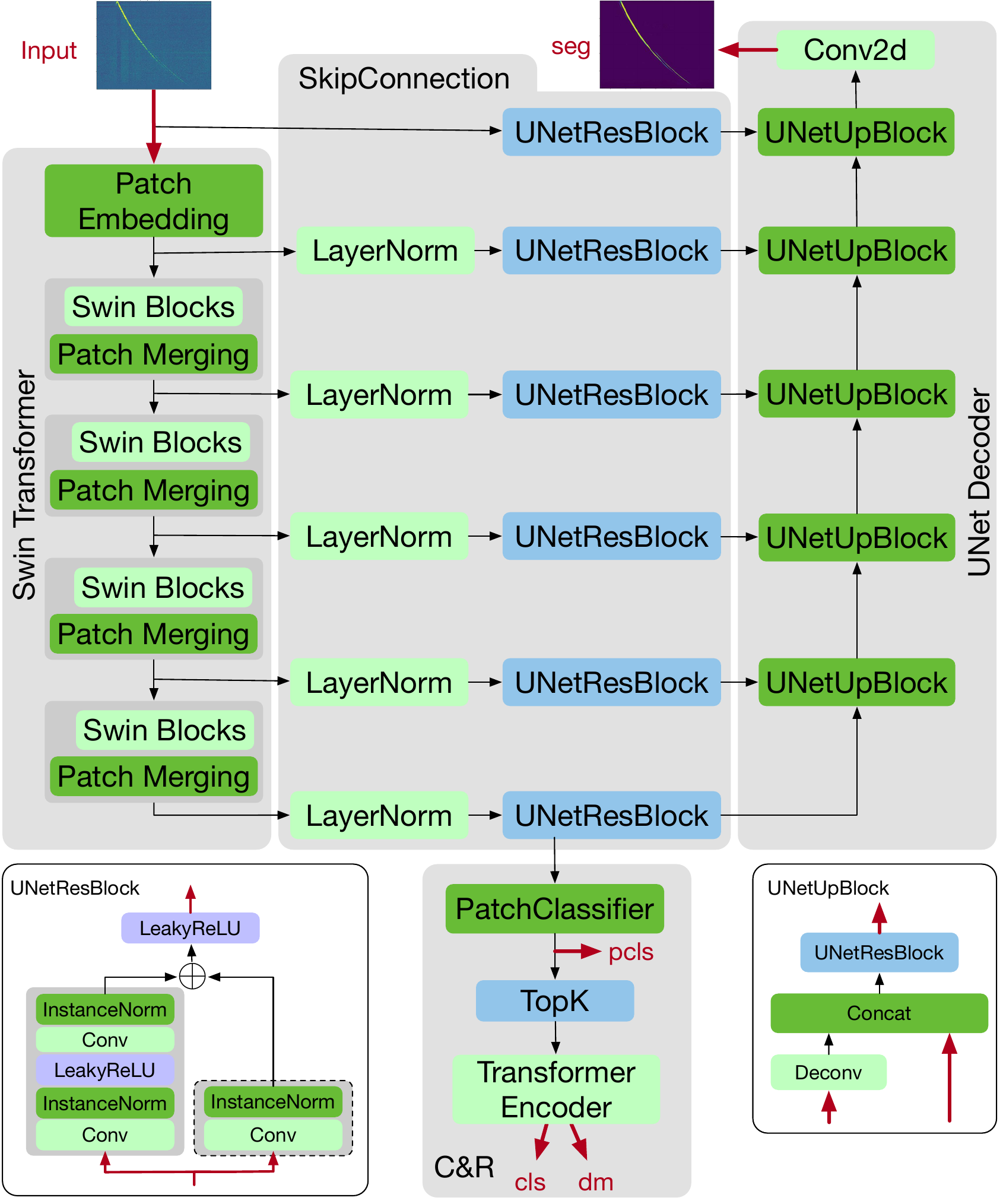}
    \caption{The \model{} model. The inputs and outputs of the model or sub-modules are indicated by the red arrows. }
    \label{fig:model}
\end{figure}

The model architecture is primarily based on U-Net \citep{ronnebergerUnetConvolutionalNetworks2015} and the Swin Transformer \citep{liuSwinTransformerHierarchical2021}, complemented by other well-established components.
U-Net employs an encoder–decoder architecture specifically designed for dense prediction tasks, such as semantic segmentation. Its defining feature is the use of skip connections that link corresponding encoder and decoder stages, allowing high-resolution spatial features to be directly fused with deep semantic representations. These skip connections mitigate information loss caused by downsampling and improve localization accuracy in the final predictions.

The Swin Transformer is a hierarchical vision transformer that redefines image modeling by partitioning images into discrete patches and mapping them into visual tokens. Unlike conventional CNNs, which rely on static convolutional kernels with inductive biases like translation invariance, the Swin Transformer employs a window-based self-attention mechanism that dynamically captures feature interactions based on input content. By incorporating a hierarchical structure similar to CNNs to generate multi-scale representations, and utilizing shifted window partitions to enable cross-window communication, it achieves global modeling capabilities with linear computational complexity. This synergy of content-adaptive weighting and multi-scale feature extraction makes it a highly effective and general-purpose backbone for diverse computer vision tasks.

The proposed architecture, \model, features a `Y'-shaped structure (Fig.~\ref{fig:model}) achieved by augmenting Swin UNETR \citep{hatamizadehSwinUNETRSwin2021} with a dedicated classification-regression (C\&R) branch.
The Swin UNETR part contains a Swin Transformer-based encoder, skip connections, and a U-Net decoder.
Conventionally, the U-Net decoder predicts a mask of FRB signals (\texttt{seg}).
The Swin Transformer-based encoder extracts multiple hierarchical feature maps from an input image.
The intermediate feature maps are linked to the \texttt{UNetUpBlock}s of the U-Net decoder by skip connections,
each of which is composed of a \texttt{LayerNorm} and a \texttt{UNetResBlock}.
The deepest feature map ($\bm{U}$), corresponding to the coarsest spatial resolution, serves as the input to the C\&R branch.
Each vector in the feature map $\bm{U}$ represents the features of a specific `patch'—a localized rectangular region—within the original input image.
We set the patch size for the Swin Transformer's input \texttt{Patch Embedding} layer to $6 \times 4$.
Accounting for the architecture's subsequent downsampling, each vector in $\bm{U}$ ultimately represents a $96 \times 64$ region of the original input image.
This dimensionsal scale has been empirically validated to align effectively with typical pulse widths.
\begin{itemize}[leftmargin=1em, topsep=6pt plus 0pt minus 2pt, itemsep=0pt, parsep=3pt minus 1pt]
    \item \texttt{PatchClassifier}: A classifier module to discriminate whether a patch contains burst signals or not.
    It takes the feature map $\bm{U}$ as input, bypasses it to the output, and simultaneously predicts the probability ($\bm{v}$) of a burst signal being present in each patch:
    $ \bm{U}, \bm{v} = \operatorname{PatchClassifier}(\bm{U}). $
    It is essential to add an output head (\texttt{pcls}) after this module to inject supervised signal such that the following \texttt{TopK}
    module get meaningful inputs.
    Supervisory labels are generated by determining the presence of a celestial signal within each $96 \times 64$ patch (\textit{see} Section~\ref{sec:labeling-rules}).
    \item \texttt{TopK}: A module to select the most probable patches that contains the FRB signal:
    $ \operatorname{TopK}(\bm{U}, \bm{v}, k) = \bm{U}[\arg\operatorname{sort}(\bm{v})[:\!k]], $
    where $k$ is a hyperparameter fixed to 128 in this work,
    $\arg\operatorname{sort}(\cdot)$ denotes the function that returns the permutation indices that sort the input array, and the square brackets indicate indexing operations.
    Since FRBs have a very low instantaneous duty cycle, most patches contain only noise or RFI.
    Analyzing only the patches most likely to contain FRB signals---e.g., those aligning with an expected dispersion sweep---suffices for reliable detection.
    The TopK module enforces this by providing a hard sparse attention mechanism to the Transformer.
    Without compromising accuracy, this module substantially reduces the number of tokens fed into the subsequent \texttt{Transformer Encoder}, thereby improving computational efficiency and reducing memory consumption.
    \item \texttt{Transformer Encoder}: A module contains 4 transformer blocks with a classification and a regression head.
    This module accept a sequence of feature vectors emitted by \texttt{TopK} as inputs and performs a sequence classification task to
    determine whether the original image contains FRB signals (\texttt{cls}).
    At the same time, if any FRB signal is detected, a regression task is conducted to estimate the DM value (\texttt{dm}).
\end{itemize}

In summary, the proposed Y-shaped architecture with contracting, expansive, and sequence transforming paths enables it to
accurately detect FRB signals, estimate DM values, and learn robust and precise segmentation masks simultaneously.
Our \model{} model, $f(\bm{x}; \bm{\theta})$, can be represented by the following formulas:
\begin{align}
    \bm{H} &= \operatorname{SwinTransformer}(\bm{x}), \nonumber \\
    \bm{h}, \bm{h}' &= \operatorname{SkipConnection}(\bm{x}, \bm{H}), \nonumber \\
    \hat{y}_{\texttt{seg}} &= \operatorname{UNetDecoder}(\bm{h}, \bm{h}'), \nonumber \\
    \bm{h}, \hat{y}_{\texttt{pcls}} &= \operatorname{PatchClassifier}(\bm{h}), \nonumber \\
    \bm{c} &= \operatorname{TopK}(\bm{h}, \operatorname{detach}( \hat{y}_{\texttt{pcls}}), k), \label{eq:topk-detach}\\
    \hat{y}_{\texttt{cls}}, \hat{y}_{\texttt{dm}} &= \operatorname{TransformerEncoder}(\bm{c}), \nonumber
\end{align}
where $\bm{H}$ represents the multiple hierarchical feature maps of the input dynamic spectrum $\bm{x}$, encoded by the Swin Transformer; $\{\bm{h}, \bm{h}'\}$ denotes the set of feature maps produced by the skip connections, where $\bm{h}$ is the deepest feature map fed into the $\operatorname{PatchClassifier}$ module; $\bm{c}$ consists of the selected feature vectors from the feature map $\bm{h}$, determined by the $\operatorname{TopK}$ operator.
The \texttt{detach} operator in Eq.~\eqref{eq:topk-detach} makes the gradient with respect to $\bm{h}$ irrelevant to
the second input of the TopK function, thereby bypassing the issue of non-differentiability.

Under the supervised learning paradigm, a dataset contains input-targets pairs are required to train the proposed model.
The target labels are intractable to obtain by manual annotations.
This is because it requires tens to hundreds of thousands annotated data to train a model and it is too expensive to do
patch- or pixel-level segmentation annotations for FRB signals that is required to train the proposed model.
As shown in Fig.~\ref{fig:model}, the targets contain four labels:
(1) a classification label (\texttt{cls}) for detection,
which indicates whether the whole input dynamic spectrum contains any FRB signal;
(2) a DM value (\texttt{dm});
(3) patches' classification labels (\texttt{pcls}), which annotate each patch containing FRB signal or not;
and (4) FRB signal segmentation mask (\texttt{seg}).
The main difficulty stem from the third and fourth labels, where both can be regarded as segmentation annotations but in different resolutions.
It is extremely difficult to annotate segmentation masks accurately at patch- or pixel-level,
especially for weak FRB signals embedded in strong noise and RFI.

On the one hand, due to the weak FRB signal, as well as the dispersion effect and strong RFI, it is difficult to distinguish the signal from the background from the original observation.
On the other hand, it is extremely difficult to accurately annotate the extremely narrow object comes from the extremely short time width of the FRB pulses (usually in the order of milliseconds) when performing pixel-level mask labeling. Because a little deviation in the direction of the time axis will lead to a great relative annotation error.

We apply FRB simulation and rule-based automatic labeling methods to overcome the above-mentioned labeling difficulties.
These approaches are detailed in the following two subsections.

\subsection{FRB Simulation Method}\label{sec:frb-simulation}
\begin{table*}[!thbp]
    \caption{Parameters used in our FRB simulator. All parameter distributions are shared across training, validation, and testing, except for the exponent of amplitude $\alpha_l$, which is shown in Table~\ref{tab:frb-simu-valid-test-params}.
    All \texttt{fitburst} parameters not listed in this table are fixed: scattering index = -4, DM index = -2, scattering timescale = 0, spectral index = 0, spectral running = -300.}%
    \label{tab:all-frb-simu-params}
    \vspace{-2.5ex}
    \begin{subtable}[t]{0.64\textwidth}%
    \centering%
    \caption{Distributions and description for simulator input parameters.}%
    \vspace{-3ex}
    \label{tab:frb-simu-params}%
    \begingroup
    \renewcommand{\arraystretch}{1.06}
    \begin{tabular}[t]{cll}%
    \toprule%
    Parameter      & Distribution             & Description \cite[\textit{see}][]{fonsecaModelingMorphologyFast2024} \\%
    \midrule
    \multirow[t]{2}{*}{$N$}  & $\operatorname{Categorical}(\bm p)$, where & \multirow[t]{2}{*}{Number of burst components.} \\%
                             & $\bm {p} =  [90, 8, 1, 0.8, 0.2]$ & $\bm {p}$ is an unnormalized distribution. \\%
    $\mathrm{DM}$ & $\mathcal{U}[50, 1400]$  & Disperse measure. \\%
    $t_{0,i}$  & $\mathcal{U}[0, 1.5)$    & Mean ToA of burst component $i$. \\%
    $\nu_{r,i}$   & $\mathcal{U}[1000,1500]$ & Pivot frequency of burst component $i$. \\%
    $\sigma_i$    & $5\times 10^{-4} + \operatorname{Exp}(500)$ &  Temporal width of burst component $i$. \\%
    $\alpha_i$    & \textit{see} Table~\ref{tab:frb-simu-valid-test-params}  & Exponent of amplitude for component $i$. \\%
    \bottomrule%
    \end{tabular}%
    \endgroup
    \end{subtable}%
    \hfill
    \begin{subtable}[t]{0.34\textwidth}%
    \setlength{\tabcolsep}{4pt}
    \centering%
    \caption{Distribution of $\alpha_i$}%
    \vspace{-3ex}
    \label{tab:frb-simu-valid-test-params}%
    \begin{tabular}[t]{clc@{}}%
    \toprule%
    Split    & Distribution & Count\\%
    \midrule%
        Train       & $\mathcal{U}[-0.75, 0]$    & $4.75\times 10^6$\\
        Valid       & $\mathcal{U}[-0.75, -0.5]$ & 1000\\%
        Test$^\dagger$ (InD)  & $\mathcal{U}[-0.75, -0.5]$ & 2000\\%
        Test (wOoD) & $\mathcal{U}[-0.9, -0.75]$ & 2000\\%
        Test (sOoD) & $\mathcal{U}[-1, -0.9]$    & 2000\\%
    \bottomrule%
    \multicolumn{3}{@{}l@{}}{%
        \parbox{0.98\linewidth}{$^\dagger$\footnotesize The test dataset consists of three subsets (see Sec.~\ref{sec:perf-on-simu-data}): in-domain (InD), weak/strong out-of-domain (wOoD/sOoD).}
    }
    \end{tabular}%
    \end{subtable}%
\end{table*}%

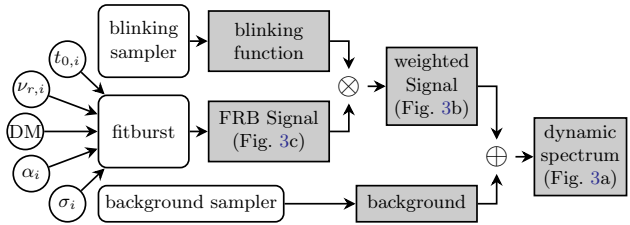
\begin{figure}[b]
    \centering
     
     \resizebox{0.99\linewidth}{!}{
     \begin{tikzpicture}[
        node distance=0.3cm and 0.5cm,
        input/.style={
            circle,
            draw=black,
            thick,
            fill=white,
            minimum size=0.6cm,
            inner sep=0pt
        },
        process/.style={
            rectangle,
            rounded corners,
            draw=black,
            thick,
            fill=white,
            minimum height=1.2cm,
            minimum width=1.5cm,
            align=center
        },
        data/.style={
            rectangle,
            draw=black,
            thick,
            fill=black!20, 
            minimum height=1cm,
            minimum width=2cm,
            align=center
        },
        op/.style={
            font=\Large 
        },
        arrow/.style={
            -{Stealth[length=2mm, width=2mm]}, 
            thick
        }
    ]

        \node[process, minimum width=1.5cm, anchor=west] (blsampler) at (0,  1.5cm) {blinking \\ sampler};
        \node[process, minimum width=1.5cm, anchor=west] (fitburst)  at (0, 0)      {fitburst};
        \node[process, minimum width=2.5cm, anchor=west, minimum height=0.6cm] (bgsampler) at (0, -1.2cm) {background sampler};

        \coordinate (xshift) at (0.2cm, 0);
        \node[input] (t0)    at ($(fitburst.west) + (-0.7cm, 1.20) + (xshift)$)  {$t_{0,i}$};
        \node[input] (nur)   at ($(fitburst.west) + (-1.3cm, 0.7)  + (xshift)$)  {$\nu_{r,i}$};
        \node[input] (dm)    at ($(fitburst.west) + (-1.4cm, 0)    + (xshift)$)  {DM};
        \node[input] (alpha) at ($(fitburst.west) + (-1.3cm, -0.7) + (xshift)$)  {$\alpha_{i}$};
        \node[input] (sigma) at ($(fitburst.west) + (-0.7cm, -1.20)+ (xshift)$)  {$\sigma_{i}$};

        %
        \node[data] (blfunc) [right=0.3cm of blsampler] {blinking \\function};
        \node[data] (frb) [right=0.3cm of fitburst]  {FRB Signal \\(Fig.~\ref{fig:simulated-frb}c)};

        \coordinate (mid_frb_bl) at ($(frb.east)!0.5!(blfunc.east)$);
        \node[op] (otimes) [right=0mm of mid_frb_bl] {$\otimes$};
        \node[data, minimum width=1.5cm] (weighted) [right=0.3cm of otimes] {weighted \\Signal \\(Fig.~\ref{fig:simulated-frb}b)};

        \coordinate (ref) at (bgsampler.north -| weighted.east);
        \node[data, anchor=north east, minimum height=0.6cm] (bg) at (ref) {background};

        \coordinate (mid_bg_weighted) at ($(bg.east)!0.4!(weighted.east)$);
        \node[op] (oplus) [right=0mm of mid_bg_weighted] {$\oplus$};

        \node[data, minimum width=0.5cm] (dynspec) [right=0.3cm of oplus] {dynamic \\ spectrum \\(Fig.~\ref{fig:simulated-frb}a)};


        \draw[arrow] (dm) -> (fitburst);
        \draw[arrow] (t0) -> (fitburst);
        \draw[arrow] (nur) -> (fitburst);
        \draw[arrow] (alpha) -> (fitburst);
        \draw[arrow] (sigma) -> (fitburst);

        \draw[arrow] (bgsampler) -> (bg);
        \draw[arrow] (fitburst) -> (frb);
        \draw[arrow] (blsampler) -> (blfunc);

        \draw[arrow] (frb.east) -| (otimes);
        \draw[arrow] (blfunc.east) -| (otimes);
        \draw[arrow] (otimes) -> (weighted.west);

        \draw[arrow] (bg.east) -| (oplus);
        \draw[arrow] (weighted.east) -| (oplus);
        \draw[arrow] (oplus) -> (dynspec.west);
    \end{tikzpicture}
    }
    \caption{Our FRB Simulator}
    \label{fig:frb-simulator}
\end{figure}

We build a FRB simulator, shown in Fig.~\ref{fig:frb-simulator},
which is based on the publicly available \texttt{fitburst} software to generate the dynamic spectra of astrophysical pulses from radio pulsars and FRBs.
This software is a powerful parametric tool for modeling the dynamic spectra of astrophysical pulses from radio pulsars and FRBs.
On the one hand, \texttt{fitburst} can precisely estimate key parameters such as DM, ToA,
and burst width by directly fitting a parametric model to the data.
This detailed characterization is crucial for understanding the underlying physics of a burst.
On the other hand, \texttt{fitburst} also utilizes both physical and heuristic parameters to generate synthetic pulse profiles via multi-component Gaussian templates, making it a versatile tool for data simulation.

\begin{figure}[tb]
    \centering
     \includegraphics[width=\linewidth]{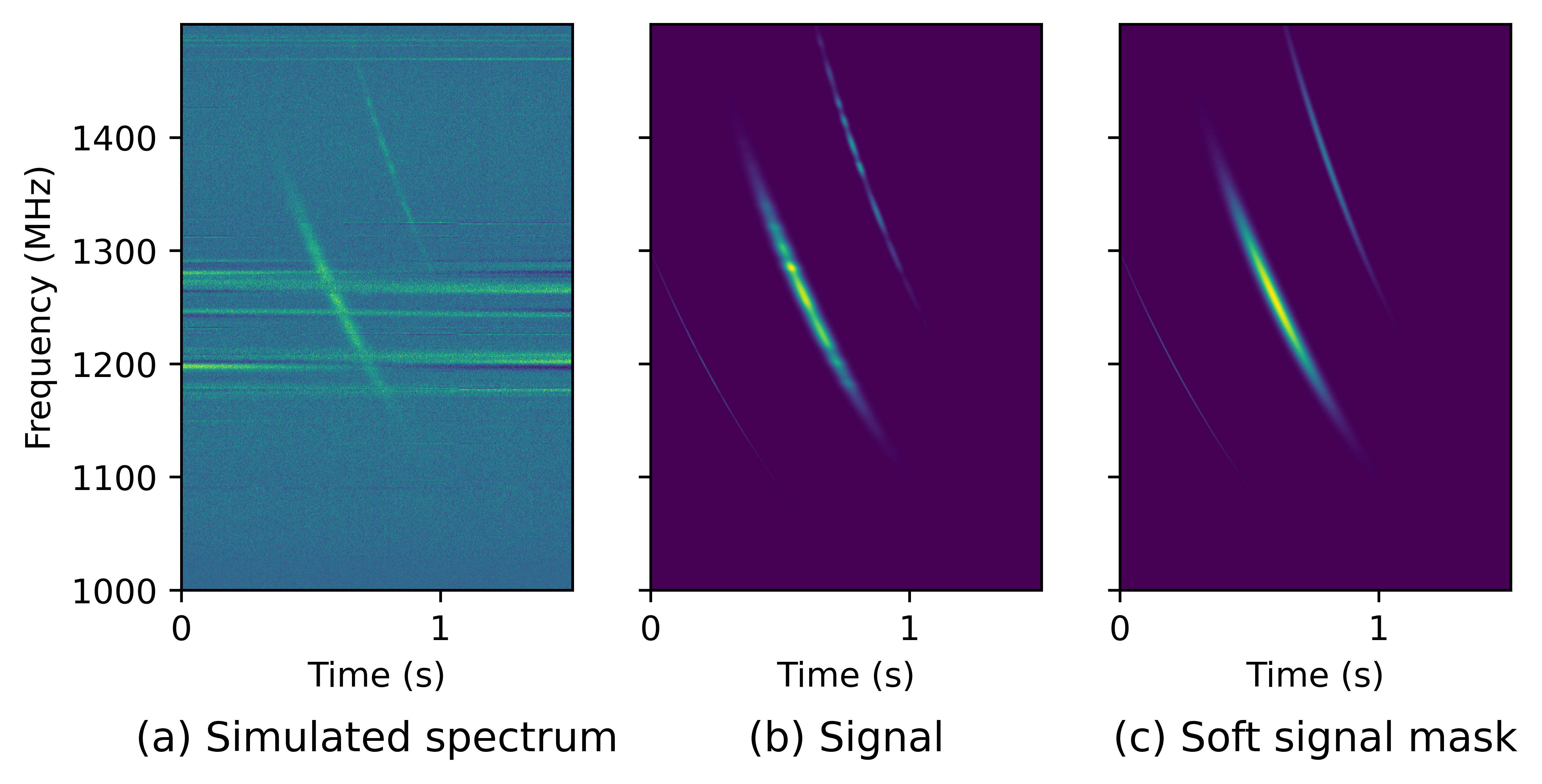}
    \caption{Simulated datum and its ground truth signal information.}
    \label{fig:simulated-frb}
\end{figure}

The \texttt{fitburst} is extended with two key components, namely, background and blinking samplers, to enable it to simulate scintillating, realistic FRBs like true observations (see Fig.~\ref{fig:simulated-frb}a).
Parameters for \texttt{fitburst} are modeled by probability distributions given in Table~\ref{tab:all-frb-simu-params}.
We first draw the number of burst components from a categorical distribution,
and then draw other parameters except $\mathrm{DM}$ for each component independently.
The $\mathrm{DM}$ parameter is shared across burst components.
The sampled parameters are fed to the \texttt{fitburst} model to generate FRB signals.
Then, the FRB signals is multiplied by a `blinking' function to amplify the signal at certain frequencies while attenuating it at others.

The `blinking' effect is modeled as a frequency-dependent function realized from a Gaussian process (GP) and constrained to the interval $[0, 2]$ via a clipping transformation. We employ a Radial Basis Function (RBF) kernel, defined as:
$ k(\nu_i, \nu_j) = A^2 \exp\left( -\frac{|\nu_i - \nu_j|^2}{2l^2} \right), $
where the amplitude $A$ and length-scale $l$ are stochastically sampled from uniform distributions such that $A \sim \mathcal{U}[0.1, 0.5]$ and $l \sim \mathcal{U}[4, 40]$.
Through multiplication, the simulated signal is re-weighted to have a scintillating effect, as shown in Fig.~\ref{fig:simulated-frb}b.
While we do not explicitly model the physical process of FRB scintillation, this re-weighting operation serves as a computationally efficient morphological approximation. To distinguish this non-physical approach from true scintillation, we refer to it as a blinking function.
In addition, it could be regarded as a data augmentation method which are widely used in deep learning
to improve generalization and enhance robustness \citep{mumuniDataAugmentationComprehensive2022}.

The \textit{background sampler} extracts samples from observation segments where no candidates were identified by PRESTO.
Unlike gravitational wave searches that rely on synthetic injections to ensure background purity \citep{George:2017pmj}, such an approach is unnecessary here due to the extreme sparsity of FRB events and the fact that deep learning models can tolerate a marginal degree of labeling noise.
To facilitate large-scale training, we prioritize computational efficiency over the exhaustive elimination of rare false negatives.
In the same interest of efficiency, rather than simulating RFI, we directly utilize real RFI obtained through sampling.
Simple sampling provides a realistic representation of the real observation without the prohibitive cost of redundant verification.

Fig.~\ref{fig:simulated-frb} shows an example generated by our built simulator.
The left panel of Fig.~\ref{fig:simulated-frb} is the simulated dynamic spectrum, which is the input of our model;
And the right panel is the corresponding FRB signals, which are used to extract the target labels.

\subsection{Rule-based Automatic Labeling}\label{sec:labeling-rules}
\begin{figure}[!htbp]
    \centering
    \includegraphics[width=0.9\linewidth]{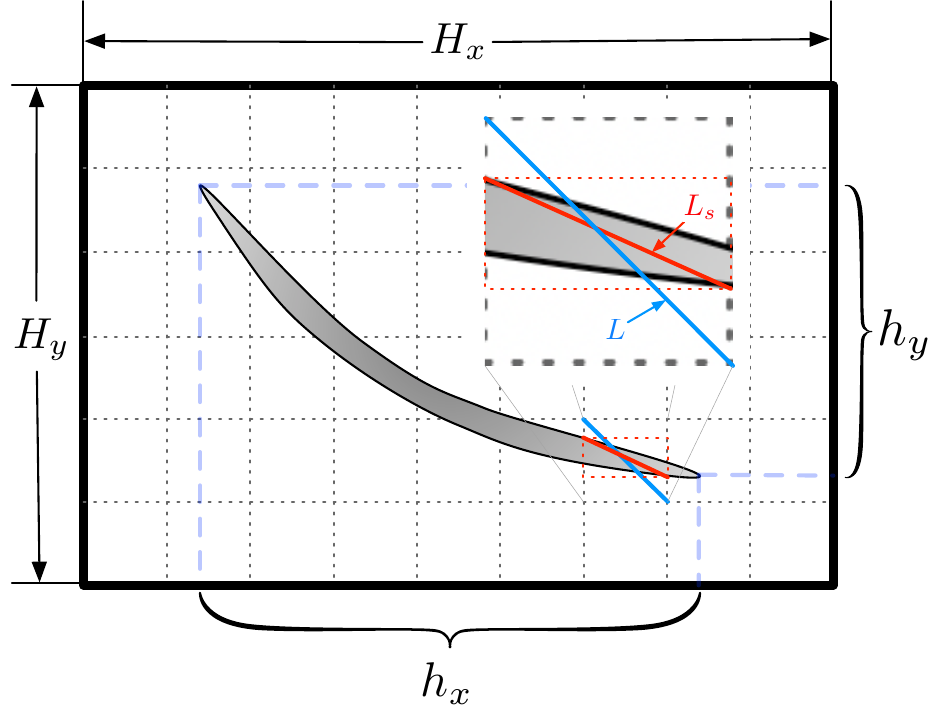}
    \caption{Key parameters used for extracting labels.}
    \label{fig:simu-label-rules}
\end{figure}

Except the trivial \texttt{dm} label, all other labels are extracted from the simulated \textit{FRB signals} (the outputs of the FRB simulator Fig.~\ref{fig:simulated-frb}b).
We employ the soft labels in model training and they are all computed according to the following formula:
\begin{equation}\label{eq:auto-soft-label}
    \text{label} = \min(\text{score} / \text{threshold}, 1),
\end{equation}
where the `threshold' is task-specific hyper-parameter and the `score' is a value representing how probable the label is to be positive.
The score can be calculated from key parameters extracted from a given FRB signal, with the meaning of these parameters illustrated in Fig.~\ref{fig:simu-label-rules}.

\paragraph{Detection labeling}

The classification score for the \texttt{cls} target is computed based on the signal's relative length,
which is obtained by projecting the FRB signal onto the time and frequency axes separately
and then calculating the ratio of the projected lengths.
The larger ratio is selected as the classification score:
$
S_{\texttt{cls}} = \max\big(\frac{h_x}{H_x}, \frac{h_y}{H_y}\big),
$

where $H_x$ and $H_y$ denote the total time span and frequency bandwidth of the simulated dynamic spectrum, and $h_x$ and $h_y$ denote the observed temporal extent and bandwidth of the simulated FRB signal in this spectrum.
The threshold in Eq.~\eqref{eq:auto-soft-label} for detection is set to be 0.25 in our experiments.

\paragraph{Patch-level segmentation labeling}
The \texttt{pcls} target serves as a segmentation target for downsampled images, a consequence of the Swin Transformer's downsampling effect. Each pixel in the downsampled image corresponds to a patch in the original image. By assigning a score to each patch in the original image, we obtain patch-level segmentation labels. These scores are computed based on the relative length of the signal overlapping with the corresponding patch.
The score for \texttt{pcls} is $ S_{\texttt{pcls}} = {L_s}/{L}$,
where $L_s$ is the diagonal length of the minimal rectangle that covers the signal (for example, the red dotted rectangle in Fig.~\ref{fig:simu-label-rules}), and $L$ is the diagonal length of a patch.
The threshold in Eq.~\eqref{eq:auto-soft-label} for patch-level segmentation is set to be 0.25.

\paragraph{Pixel-level segmentation labeling}

The segmentation score for the \texttt{seg} target is derived from normalized signals, defined as the normalized sum of all pulse components.
Each pixel's segmentation score corresponds to its normalized intensity.
The threshold in Eq.~\eqref{eq:auto-soft-label} for pixel-level segmentation is set to be 0.4.

In summary, we propose a simple yet effective method to assign belief values to three sub-tasks---\texttt{cls}, \texttt{pcls}, and \texttt{seg}---on simulated data.
These belief values are used as soft labels to supervise network training.

\subsection{Training and Inference}\label{sec:training-inference-method}
\paragraph{Training Settings}
The observations of a radio telescope are time-frequency intensity measurements of radio emissions, called dynamic spectra, typically visualized as two-dimensional plots
(see Fig.~\ref{fig:simulated-frb}a).
For a given input  dynamic spectrum $\bm{x}$, the model prediction $\hat{\bm{y}}$ is obtained through forward propagation,
while the corresponding ground truth label $\bm{y}$ is generated using our rule-based automatic labeling approach.
Both the ground truth $\bm{y}$ and model prediction $\hat{\bm{y}}$ are 4-dimensional tuples,
where each element corresponds to one of the four target tasks:
\begin{align*}
    \bm{y} &= (y_\texttt{cls}, y_\texttt{dm}, y_\texttt{pcls}, y_\texttt{seg}), \\
    \hat{\bm{y}} &= f(\bm{x}; \bm\theta) =(\hat y_\texttt{cls}, \hat y_\texttt{dm}, \hat y_\texttt{pcls}, \hat y_\texttt{seg}),  \label{eq:model-outputs}
\end{align*}
where $f$ our model and $\bm{\theta}$ is its trainable parameters;
$y_t$ and $\hat{y}_t$ are the task-specific label and the model's output for that specific task $t$.
Here, $t \in T = \{\texttt{cls}, \texttt{dm}, \texttt{pcls}, \texttt{seg}\}$, the set of tasks.

We optimize the model parameters $\bm{\theta}$
to minimize the discrepancy between the model's predictions and the ground truth labels.
This is achieved by minimizing the empirical loss
\begin{align}
\mathcal{L}(\bm\theta) &= \frac{1}{|\mathcal{D}|}\sum_{(\bm{x}, \bm{y}) \in \mathcal{D}} \mathcal{L'}(\bm{\theta}; \bm{x}, \bm{y}),  \nonumber
\end{align}
where $\mathcal{D}$ is the training dataset and $\mathcal{L'}(\bm{\theta}; \bm{x}, \bm{y})$ represents the per-instance loss, defined as a weighted sum of task-specific losses:
\begin{align}
    \mathcal{L'}(\bm{\theta}; \bm{x}, \bm{y}) &=
        \sum_{t\in T}\lambda_t \mathcal{L}_{t}(f_{t}(\bm{x}; \bm{\theta}), y_{t})  \nonumber
\end{align}
where $\lambda_t$, $\mathcal{L}_t(\cdot, \cdot)$, and $f_{t}(\bm\theta; \bm{x}) = \hat{y}_t$
is the weight, the task-specific loss function, and the model's output for task $t$, respectively.
The per-task losses are defined as follows:
\begin{align*}
    \mathcal{L}_\texttt{cls}(\hat{y}, y) &= \operatorname{CrossEntropyLoss}(\hat{y}, y) \nonumber \\
    &= -y\log \hat{y} - (1-y) \log(1-\hat{y}), \\
    \mathcal{L}_\texttt{dm}(\hat{y}, y) &= \operatorname{MSELoss}(\hat{y}, y) = (\hat{y} - y)^2, \\
    \mathcal{L}_\texttt{pcls}(\hat{y}, y) &= \operatorname{FocalLoss(\hat{y}, y)}, \\
    \mathcal{L}_\texttt{seg}(\hat{y}, y) &= \operatorname{FocalLoss(\hat{y}, y)} = -\alpha (1 - p_t)^\gamma \log p_t,
\end{align*}
where $p_t$ is the probability of the correct classification \citep{linFocalLossDense2020}.

To ensure robust generalization to real observational data, our model requires training on an extensive simulated dataset that comprehensively captures the statistical properties and variability of the target domain.
Through empirical validation, we determined that 4.75 million simulated instances provide sufficient coverage for effective model training.
While complete storage of all these simulated data would exceed 355 TB, we mitigated this requirement through on-the-fly simulation and immediate consuming the simulated results for training.
It takes 21 days using an 43-core Xeon Platinum 8358 server with 8 NVIDIA A40 GPUs to complete the training.

We utilized the AdamW optimizer \citep{loshchilovDecoupledWeightDecay2019} with a cosine decay learning rate schedule.
The learning rate was initialized at $1\times10^{-3}$ with a minimum of $1\times 10 ^ {-5}$, incorporating a linear warm-up phase over the first 30,000 training steps.
The learning rate was selected based on performance on a validation set of 2,000 samples, after training on a small simulated dataset of 300,000 examples.
The validation set consists of 1000 negative examples randomly drawn from the CRAFTS project \citep{li2018fast} and
1000 positive instances simulated using parameters in Table~\ref{tab:frb-simu-params} and Table~\ref{tab:frb-simu-valid-test-params}.

To accommodate memory limitations, we adopt a training configuration with per-GPU batch size of 1 and 10-step gradient accumulation. Through distributed data parallel training on 8 GPUs, this achieves an effective batch size of 80.

\paragraph{Inference Pipeline}

Our model ingests downsampled $7680\times 2048$ matrices (1.5s dynamic spectra;
$\Delta t = 19.661$ms; $\Delta \nu = 0.2441$MHz)---reduced-resolution versions of FAST's native observations through independent time-frequency scaling.
By design, our model can only accept fixed resolution inputs.
As a consequence, the observed dynamic spectra must be converted to match our model's input specification in order to perform accurate inference.

Given an input spectrum $S(t,\nu)$, we generate samples $\{S_i\}$
via:
\begin{equation*} 
    S_i = \mathcal{R}(W_i \odot S),
\end{equation*}
where $W_i$ is the sliding window function ($1/3$ overlap) and $\mathcal{R}$ is the downsampling operator to target dimensions $7680\times 2048$.
Then, all the chunks are feed to the model independently for inference.
Because a FRB signal can be observed in multiple segments, the model can emit several positive predictions for a given spectrum.
A burst is detected if at least one of the segments $\{S_i\}$ are classified as positive.

Optional post-processing procedures can be performed base on the neural network's prediction.
The first one is pulse detection via instance segmentation.
The prediction $y_\texttt{seg}$ signify each pixel belongs to a FRB signal or not.
However, it does not show whether they are belonging to the same pulse or to different pulses.
By instance segmentation, we can distinguish multiple pulses occur in the same observation chunk.

The second one is parameter estimation via linear regression.
Even though, our model can predict the DM value end-to-end from the raw dynamic spectra, it has relatively high error.
We can get a better DM value by fitting Eq.~\eqref{eq:time-freq-dm}
using the points of each pulse extracted from the first optional post-processing procedure.
Firstly, denoting $k=k_\text{DM} \cdot \mathrm{DM}$, $b=-k\nu_0^{-2}+t_0$, and $x=\nu^{-2}$, this equation can be rewrite to $t = kx + b$.
Secondly, the estimated $\hat{k}$ and $\hat{b}$ can be obtained by linear regression.
And finally, the DM value and the ToA ($t_0$) can be calculated by $\mathrm{DM} = \hat{k}/k_{\mathrm{DM}}$ and
$t_0 = \hat{b} + \hat{k} \nu_0^{-2}$, respectively.
Note that we propose only a simple and fast method for estimating the DM value,
without pursuing to obtain the DM that yields the optimal signal-to-noise ratio \citep{zackayAccurateEfficientAlgorithm2017}
or the one that is structurally  optimal in any specific sense \citep{seymourDM_phaseAlgorithmCorrecting2019,Lin:2022qzf}.

\section{Experimental Results and Analysis}\label{sec:experiment-and-analysis}
In this section, we comprehensively evaluate the detection, parameter estimation, and segmentation capabilities of our model using the ``FAST dataset for Fast Radio bursts EXploration'' (FAST-FREX) \citep{guoFASTFREXFASTDataset2024,guoAcceleratingFRBSearch2024} and a complementary simulation dataset.
The FAST-FREX dataset, derived from FAST observations, addresses the scarcity of high-quality FRB data for machine learning. It contains 600 positive samples from FRB20121102, FRB20180301, and FRB20201124, and 1000 negative samples of RFI and noise. Each positive sample spans $\sim$6.04\,s, contains one FRB event, and is accompanied by a CSV file with parameters such as DM and ToA; negative samples lack such metadata. Stored in FITS format, the data have sampling rates of \(98.304\,\mu\mathrm{s}\) for FRB20121102 and \(49.152\,\mu\mathrm{s}\) for the others, with 4096 frequency channels covering 1--1.5\,GHz at 0.122\,MHz resolution and a single polarization. This balanced dataset provides a robust basis for developing efficient machine learning methods beyond traditional search techniques.

\subsection{FRB Detection Performance}%
\begin{table*}[!tbp]%
\centering%
\caption{Detection results on the FAST-FREX dataset.
\model/m refers to a model with $1/m$ the standard resolution, corresponding to an $m$-fold increase in $\Delta t$ and $\Delta \nu$ relative to the baseline.
``TP'' = true positives, ``FP'' = false positives, ``FN'' = false negatives. PRESTO is evaluated under three S/N thresholds and its statistics are from \citet{zhangDRAFTSDeepLearningbased2024}.
Heimdall statistics are from \citet{guoAcceleratingFRBSearch2024}.
Duplicates indicate redundant detections of the same event.
The \textit{time} column reports the average inference time over the positive subset, including neural network evaluation as well as pre- and post-processing.
Optional post-processing procedures described in section~\ref{sec:training-inference-method} are disabled for our model.
Runtime is measured on a Linux machine equipped with an Intel i7-12700KF CPU and a GeForce RTX 3080 Ti GPU.
}%
\label{tab:frex-results}%
\begin{tabular}{ccccccccccc}%
\toprule%
Method & Threshold & TP($\uparrow$) & FP($\downarrow$) & FN($\downarrow$) & Duplicates($\downarrow$) & Precision($\uparrow$) & Recall($\uparrow$) & F1($\uparrow$) & time($\downarrow$) \\%
\midrule%
\multirow[t]{3}{*}{PRESTO}%
          & $\mathrm{S/N}=3$ & 520 & 10,663,950 & 80 & 43,044 & 0.049\permil  & 86.7\%        & 0.098\permil & - \\%
          & $\mathrm{S/N}=5$ & 513 & 17,406     & 87 & 40,818 & 2.8\%          & 85.5\%          & 5.4\%     & - \\%
          & $\mathrm{S/N}=7$ & 477 & 4,488      & 123& 25,402 & 9.6\%          & 79.5\%          & 17.1\%    & - \\%
Heimdall  & -                & 489 & 5,854      & 36 & -      & 7.7\%          & 81.5\%          & 14.1\%    & - \\%
\midrule%
RaSPDAM   & -                & 501 & 14         & 67 & -      & 97.3\%         & 83.5\%          & 89.9\%          & 8.37 \\%
DRAFTS    & 0.5              & 580 & 23         & 20 & -      & 96.2\%         & 96.7\%          & 96.4\%          & 14.19 \\%
\model/4  & 0.5              & 576 & 3          & 24 & -      & 99.5\%         & 96.0\%          & 97.7\%          & \textbf{4.34} \\%
\model/2  & 0.5              & 581 & 4          & 19 & -      & 99.3\%         & \textbf{96.8\%} & \textbf{98.1\%} & 4.97 \\%
\model    & 0.5              & 574 & 0          & 26 & -      & \textbf{100\%} & 95.7\%          & 97.8\%          & 6.07 \\%
\bottomrule%
\end{tabular}%
\end{table*}%

We evaluate our model's detection performance on the FAST-FREX dataset. Table~\ref{tab:frex-results} demonstrates that traditional FRB search pipelines, such as PRESTO and Heimdall, produce false positives at rates several orders of magnitude higher than modern deep learning-based methods. In contrast, AI-based approaches yield less than two dozen false positives on the same test set. Remarkably, our model produces zero false positives, which, while indicative of its strong discriminative ability, may also reflect limitations in the dataset's coverage of edge cases or noise patterns. Furthermore, the large number of duplicate detections reported by PRESTO suggests a reduced capacity for resolving individual pulses. Overall, these results underscore the effectiveness of deep learning models in suppressing RFI and reliably identifying genuine FRB signals.

Among AI-based baseline models, our model achieves the highest F1 score, surpassing the strongest baseline, DRAFTS, by 1.4\%,
with a perfect precision and a slightly lower recall (–1\%), eliminating all false positives.
Our model outperforms RaSPDAM by a substantial margin in both precision (+2.7\%) and recall (+12.2\%).
Notably, our framework also supports pixel-wise segmentation, which is not available in the baselines.

Lowering the input resolution has a negligible impact on search metrics, with the F1-score varying within $\pm 0.3\%$.
Although some variants improve recall by up to 1.1\% and accelerate inference,
they compromise the spatial resolution of the output semantic segmentation.
Unless otherwise specified, all subsequent analyses are conducted using the standard model rather than its reduced-resolution variants.

\begin{table}[htbp]
    \centering
    \caption{Recall across Different S/N Ranges.}
    \label{tab:recall-vs-snr}
    \begin{tabular}{r@{\hspace{0.5em}} r@{\hspace{1.5em}} r@{\hspace{2em}} r@{\hspace{1em}} r}
        \toprule
        \multicolumn{1}{c}{S/N range} & \multicolumn{1}{c}{Correct} & \multicolumn{1}{c}{Incorrect} & \multicolumn{1}{c}{Total} & \multicolumn{1}{c}{Recall} \\
        \midrule
        $(3, 5]$       & 38  & 3  & 41  & 92.7\% \\
        $(5, 7]$       & 85  & 14 & 99  & 85.8\% \\
        $(7, 9]$       & 105 & 9  & 114 & 92.1\% \\
        $(9, +\infty)$ & 346 & 0  & 346 & 100\%  \\
        \bottomrule
    \end{tabular}
\end{table}

We evaluated the model's detection performance across varying Signal-to-Noise Ratios (SNRs), as summarized in Table~\ref{tab:recall-vs-snr}. SNRs were determined using the \texttt{riptide} library \citep{morelloOptimalPeriodicitySearching2020} based on parameters from the FAST-FREX dataset.
Precision is omitted as no false positives occurred.

Counterintuitively, recall does not increase monotonically with SNR; performance degradations appear in certain high-SNR intervals. Post-hoc analysis of the 26 missed detections revealed that while the model localized the signal in 88.5\% (23/26) of cases, the predicted masks exhibited lower activations or higher noise levels compared to correctly classified samples. We hypothesized that the model's sensitivity is governed by local saliency rather than integrated global SNR, meaning signals with high energy but low local contrast (e.g., highly dispersed or broad-band signals) may be missed.

To test this, we partitioned each spectrum into 16 sub-bands for independent SNR calculation. Supporting our hypothesis, only 7\% of sub-bands in missed samples had an $\text{SNR} > 3$, compared to 34\% in correctly classified samples. These results confirm that our model prioritizes locally salient features, offering a detection characteristic that complements traditional boxcar-matching pipelines (e.g., Heimdall and PRESTO) which primarily rely on global SNR.

We further applied our model to blind searches for FRBs using several days of CRAFTS data \citep{li2018fast}, as detailed in Section~\ref{sec:blind-search-application}.
The model consistently achieved a very low false positive rate (below 0.5\% across all observation days), significantly reducing the burden of manual verification.
Furthermore, the practical efficacy of the pipeline was confirmed by the successful discovery of two pulsars, maintaining a remarkably low false discovery rate of 2.8\%.

To evaluate the impact of RFI on model performance, we conducted a manual post-hoc inspection of the inference results.
From a random sample of 200 true negatives selected from 1,000 negative instances, 181 (90.5\%) exhibited identifiable RFI features,
demonstrating the pipeline's robustness in filtering non-astrophysical noise.
Regarding false alarms, as no file-level false positives occurred in FAST-FREX, we analyzed the results at the segment level.
Out of 3,600 segments, all 7 misclassified instances (100\%) were caused by RFI.
To supplement this small sample, we extended our analysis to 356 false alarms identified during a large-scale blind search, finding that 345 (96.9\%) were RFI-induced.
These results underscore that while our model significantly suppresses interference, residual RFI remains the primary driver of false alarms.

These results collectively underscore the model's exceptional generalization capabilities.
The successful transition from synthetic training environments to the complexities of real-world FAST observations validates our underlying methodology.
By achieving superior performance on the FAST-FREX benchmark and facilitating the discovery of real signals with minimal manual overhead, the model demonstrates its readiness for large-scale, automated FRB searches in the next generation of radio surveys.

\subsection{Runtime Performance}

\begin{table}[!htb]
    \centering
    \caption{
    Average runtime breakdown (in seconds) of different models on positive samples from the FAST-FREX dataset.
    Inference is considered real-time if completed within 6 seconds.
    \textbf{pre.}: preprocessing; \textbf{net}: neural network inference; \textbf{post.}: postprocessing;
    \textbf{infer}: total inference time (pre. + net + post.);
    \textbf{IO}: data loading/saving time; \textbf{Total}: sum of inference and IO times.
    }
    \label{tab:runtime-comparision}
    \begin{tabular}{cd{2.2}ccd{2.2}cd{2.2}}
    \toprule
    Model   & \multicolumn{1}{c}{pre.} & net & post. & \multicolumn{1}{c}{infer} & IO & \multicolumn{1}{c}{Total}  \\
    \midrule
    RaSPDAM & 4.15  & 4.20 & 0.02  & 8.37  & 0.38 & 8.75  \\
    DRAFTS  & 14.04 & 0.11 & 0.04  & 14.19 & 5.51 & 19.71 \\
    \model  & 3.77  & 2.30 & 0.00  & 6.07  & 2.62 & 8.69 \\
    \bottomrule
    \end{tabular}
\end{table}

In terms of runtime efficiency, our model is the fastest among the strong baselines, achieving an inference time of 6.07 seconds---approaching real-time performance---even with a trivial implementation.
By trivial implementation, we mean that preprocessing, neural network inference, and post-processing are performed sequentially, with no optimizations applied to overlap these operations.
This measurement excludes I/O time, which can be effectively hidden through overlapping computation with I/O operations.
We do not report timing statistics for traditional methods in Table~\ref{tab:frex-results} for two reasons. First, prior work has already demonstrated the superior efficiency of AI-based models over traditional approaches~\citep{zhangDRAFTSDeepLearningbased2024}. Second, tools such as PRESTO and Heimdall do not clearly separate preprocessing, inference, and post-processing stages, making it difficult to isolate and measure their inference time accurately.

Table~\ref{tab:runtime-comparision} presents a detailed breakdown of runtime components for RaSPDAM, DRAFTS, and our proposed \model{} model.
When including I/O overhead, our model and RaSPDAM exhibit comparable total runtimes, because our method spend more time on preparing and saving the segmentation masks.
Both \model{} and RaSPDAM exhibit a balanced runtime distribution between neural network inference and other operations. Consequently, they can be further optimized for true real-time processing using standard engineering techniques, such as parallelizing I/O, preprocessing, and post-processing.
In contrast, DRAFTS is less amenable to such optimizations. Its dual-model architecture and heavy reliance on computationally intensive de-dispersion steps (preprocessing time 128$\times$ longer than network inference) introduce significant bottlenecks, limiting GPU utilization and overall throughput.
All timing statistics are measured on a consumer-grade PC equipped with a single RTX 3080 Ti GPU.

In summary, our model is efficient enough to enable real-time FRB or pulsar searches using only a single consumer-grade GPU.

\subsection{Parameter Estimation and Segmentation Quality}
Since the FAST-FREX dataset does not provide pixel-level segmentation masks, we evaluate the model’s segmentation performance indirectly through a downstream parameter estimation task. Specifically, we reformulate the segmentation assessment as an evaluation of the accuracy of DM and ToA estimates derived from the predicted segmentation masks. Low-quality masks typically fail to yield reliable time-frequency coordinates, resulting in poor curve fits, and thus serve as an effective proxy for segmentation reliability. To this end, we extract time-frequency coordinates from the predicted masks and fit the dispersion relation defined in Eq.~\eqref{eq:time-freq-dm} to obtain DM and ToA estimates. The accuracy of these estimates is then used to assess the quality of the segmentation output.

\begin{table}[!htbp]
\centering
\caption{Errors in DM and ToA parameter estimates. ``Direct'' refers to direct neural outputs from the \texttt{dm} head, while ``Fitted'' uses curve fitting from segmented masks. Values are in $\mathrm{pc\cdot cm^{-3}}$ for DM and milliseconds for ToA.}
\label{tab:error-params}
\begin{tabular}{rrd{2.1}@{${} \pm {}$}lr@{${} \pm {}$}l}
\toprule
\multicolumn{1}{c}{Type} & \multicolumn{1}{c}{Source} & \multicolumn{2}{c}{$\varepsilon_{\mathrm{DM}}$} & \multicolumn{2}{c}{
$\varepsilon_{\mathrm{ToA}}$} \\
\midrule
\multirow[t]{4}{*}{Direct}
& FRB20121102 & -1.3 & 16.7 & \multicolumn{2}{c}{--} \\
& FRB20180301 & -0.3 & 26.3 & \multicolumn{2}{c}{--} \\
& FRB20201124 & -8.1 & 18.5 & \multicolumn{2}{c}{--} \\
& Total       &  0.7 & 17.6 & \multicolumn{2}{c}{--} \\
\midrule
\multirow[t]{4}{*}{Fitted}
& FRB20121102 & 0.4 & 8.0 & 1.3 & 9.5 \\
& FRB20180301 & 6.5 & 7.3 & 11.9 & 14.1 \\
& FRB20201124 & 3.2 & 2.9 & –1.5 & 3.7 \\
& Total       & 1.1 & 7.3 & –0.8 & 8.8 \\
\bottomrule
\end{tabular}
\end{table}

\begin{figure}[!htbp]
    \centering
    \includegraphics[width=0.99\linewidth]{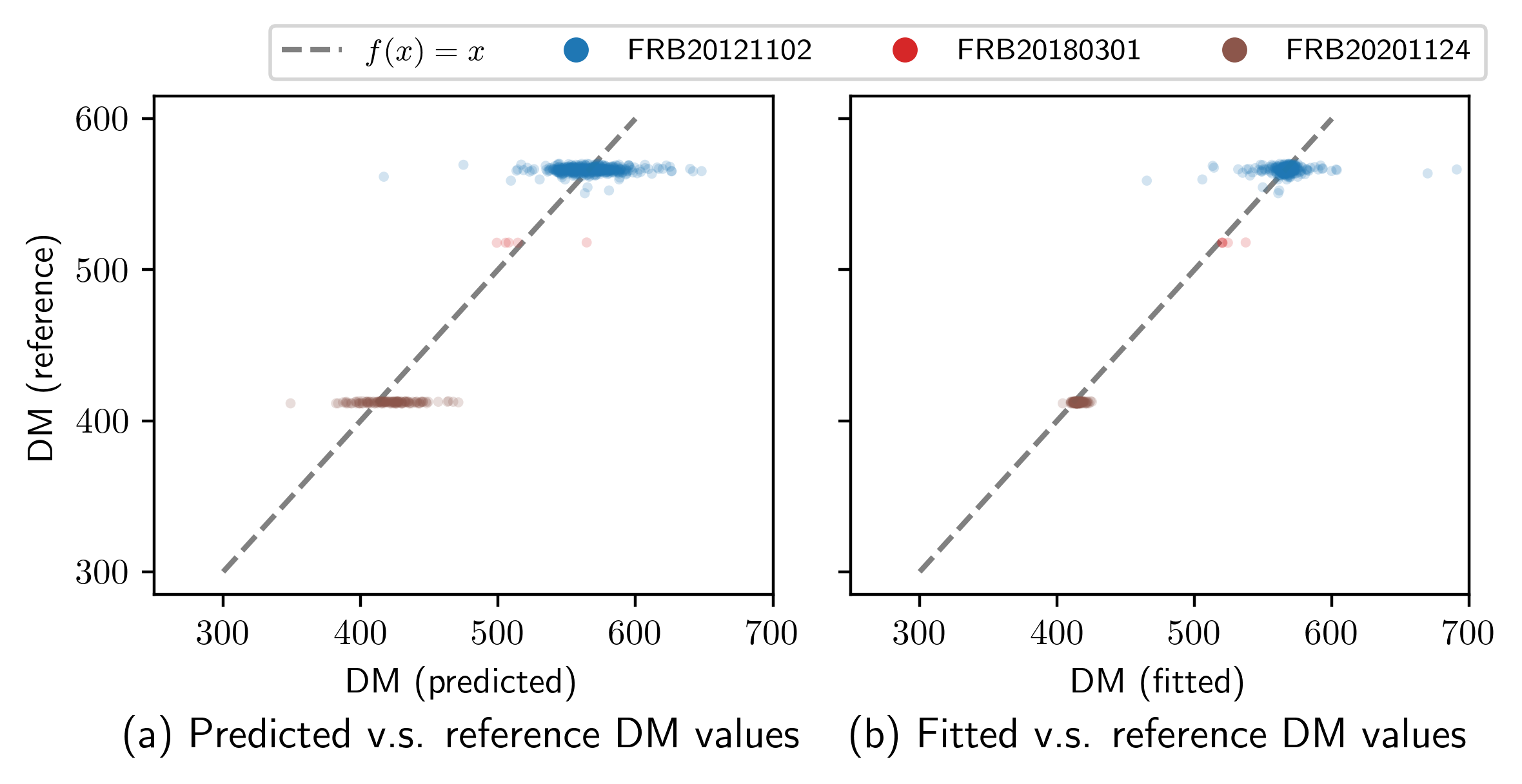}
    \caption{Predicted vs. reference DM values. The gray dashed line represents $f(x) = x$, indicating perfect predictions.}
    \label{fig:dm-values}
\end{figure}

\begin{figure}[!htbp]
    \centering
    \includegraphics[width=0.99\linewidth]{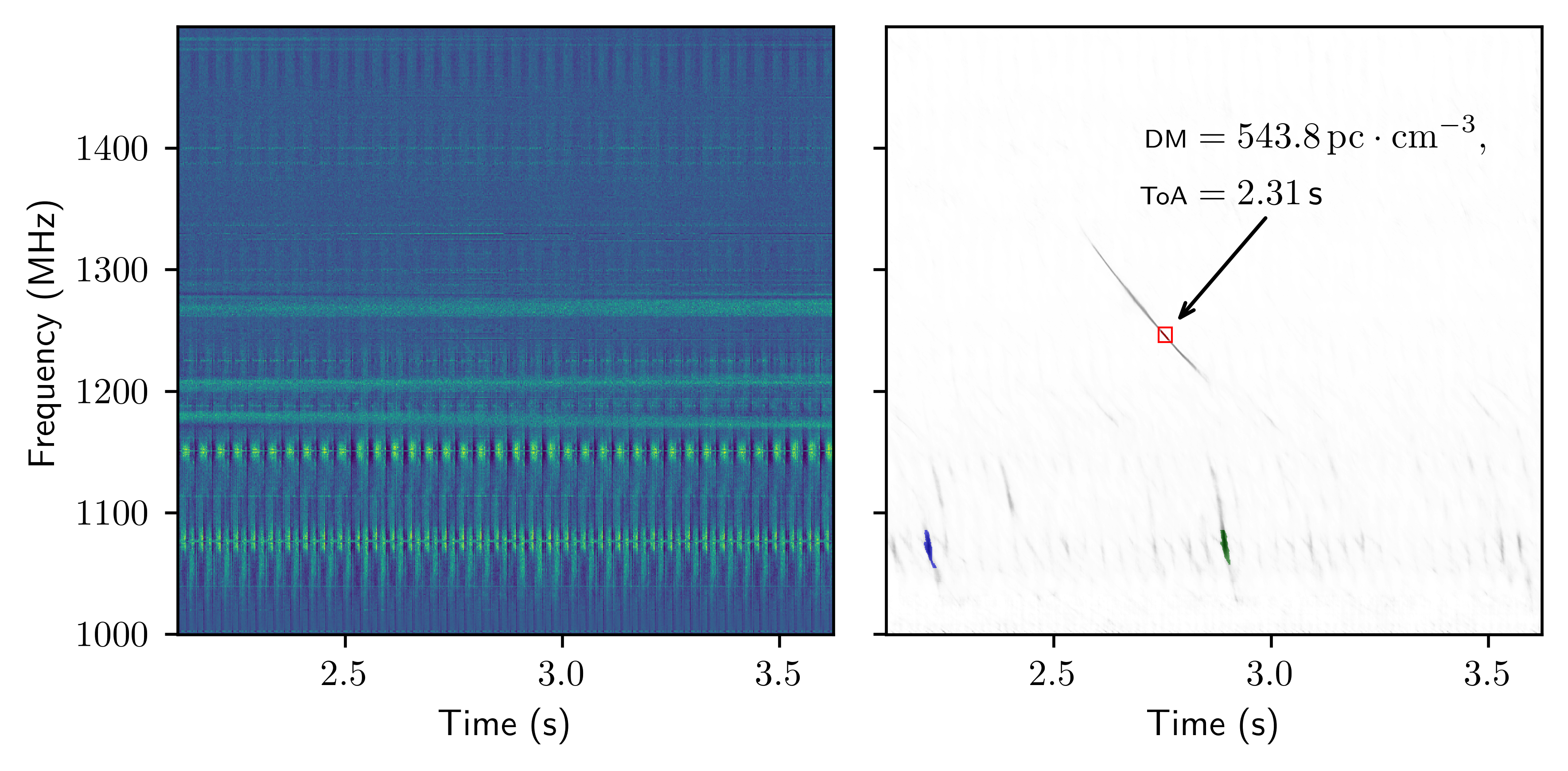}
    \caption{An example of a poorly fitted parameters, illustrating the challenge of weak segmentation.
    Partial detection of the true pulse (in the red rectangle bounding box) and confusion with strong RFI led the model to incorrectly identify false positives as the main signal, resulting in significant error.
    Fitted DM = $73.9\, \text{pc}\cdot \text{cm}^{-3}$, ToA = 2.7644\,s;
    Ground truth DM = $564.3\, \text{pc}\cdot \text{cm}^{-3}$, ToA = 2.2867\,s.}
    \label{fig:bad_case_dm_outlier}
\end{figure}

There are two ways to estimate the DM values.
The first is the direct output from the model's \texttt{dm} task, and the second is obtained by fitting Eq.~\eqref{eq:time-freq-dm}, as described in Section~\ref{sec:training-inference-method}.
Fig.~\ref{fig:dm-values} confirms that most DM predictions, both direct and fitted, align closely with ground-truth values across all three FRB sources. Outliers are rare but mostly originate from FRB20121102.
Fig.~\ref{fig:bad_case_dm_outlier} shows a bad case that the DM value is poorly fitted due to bad prediction
of signal segmentation.
In this case, the pulse located near the center of the dynamic spectrum is not fully recognized by the model.
Although a relatively long pulse is visually apparent, only a small portion---highlighted by the red bounding box---is detected,
as it fails to surpass the 0.5 decision threshold.
Fitting parameters to this partial detection yields a DM error of
$20.5 \text{pc}\cdot \text{cm}^{-3}$.
Due to the presence of strong and morphologically complex RFI, the model incorrectly predicts two additional pulses (marked in green and blue in the lower part of the later subplots).
Since these false positives contain more pixels, the post-processing algorithm mistakenly identifies them as the main signal components for DM estimation, ultimately leading to an erroneous result.

Quantitatively, fitted DM estimates exhibit substantially lower variance ($\pm7.3$ vs.\ $\pm17.6$) and smaller absolute error (\textit{see}~Table~\ref{tab:error-params}), making them more reliable than direct predictions. The ToA predictions, available only through fitting, yield an overall error of $-0.8 \pm 8.8$ ms, confirming the viability of the segmentation-based approach for time localization.

These quantitative evaluations demonstrate that our model can effectively isolate astrophysical signals from various forms of complex RFI, enabling more accurate morphological analysis and extraction of physical parameters.

\subsection{Performance on Simulated Data} \label{sec:perf-on-simu-data}
We further simulated 6,000 test examples using the parameters presented in Table~\ref{tab:frb-simu-params} and Table~\ref{tab:frb-simu-valid-test-params}
to evaluate our model, especially on its signal segmentation and DM value prediction abilities.
The simulated test data are divided into three distinct subsets, each containing 2,000 examples:
(1) In-domain (\textit{InD}): Generated using identical parameter ranges as the validation dataset, which is a subset of the training dataset;
(2) Weak out-of-domain (\textit{wOoD}): Featuring moderately shifted amplitude distributions relative to the validation set; and
(3) Strong out-of-domain (\textit{sOoD}): Simulated with significantly different amplitude distributions.

The evaluation metrics on the simulated test data is presented in Table~\ref{tab:sim_results}.
Note the simulated test data are all positive examples, and hence precision is not reported.
The DM errors are only computed on the true positive examples, with outliers excluded.
\begingroup
\addtolength{\tabcolsep}{-0.3em}
\begin{table}[!htbp]
\centering
\caption{Performance on simulated test data}
\begin{tabular}{lccccc}
\toprule
\multicolumn{1}{c}{Metric} & InD    & wOoD   & sOoD   & OoD    & Total \\
\midrule
$\varepsilon_{\mathrm{DM}}$ {\small(direct)}
            & ${\scriptstyle -2.5}{\scriptscriptstyle\pm 13.3}$ &  ${\scriptstyle -5.6}{\scriptscriptstyle\pm 18.6}$  & ${\scriptstyle -8.8}{\scriptscriptstyle\pm 21.0}$
            & ${\scriptstyle -7.2}{\scriptscriptstyle\pm 22.4}$ & ${\scriptstyle -5.6}{\scriptscriptstyle\pm 24.3}$ \\
{\small Recall}      & 98.2\% & 95.1\% & 93.4\% & 94.3\% & 95.6\% \\
{\small mIoU}        & 96.7\% & 95.3\% & 90.7\% & 93.2\% & 94.5\% \\
{\small Segm. Rec.}  & 97.6\% & 96.8\% & 93.6\% & 95.3\% & 96.2\% \\
{\small Segm. Prec.} & 99.0\% & 98.5\% & 96.7\% & 97.7\% & 98.2\% \\
\bottomrule
\end{tabular}
\label{tab:sim_results}
\end{table}
\endgroup
It reveals several important patterns in our model's performance.
The DM estimation accuracy is robust and similar to that on the FAST-FREX datasets prestented in Table~\ref{tab:error-params}.
Recall declines gradually under distribution shifts, indicating expected yet controlled degradation. Even on the most challenging out-of-domain split, our model achieves a recall of 93.4\%.
Segmentation metrics show strong resilience to distribution shifts, which suggests our architecture effectively preserves spatial localization capabilities under distribution shifts.
In conclusion, our model achieves remarkable robustness to moderate distribution shifts
and practical utility even under strong distribution shifts.

\subsection{Ablation Studies}
\paragraph{Architecture variants}
To identify the key components that effectively model dynamic spectra observed by radio telescopes, we developed and evaluated several alternative neural network architectures.
Given the intractability of exhaustively exploring all possible network configurations,
we focused on a systematic selection of representative models that incorporate distinct structural and computational features.

\begin{figure}[!htbp]
    \centering
    \includegraphics[width=0.4\linewidth]{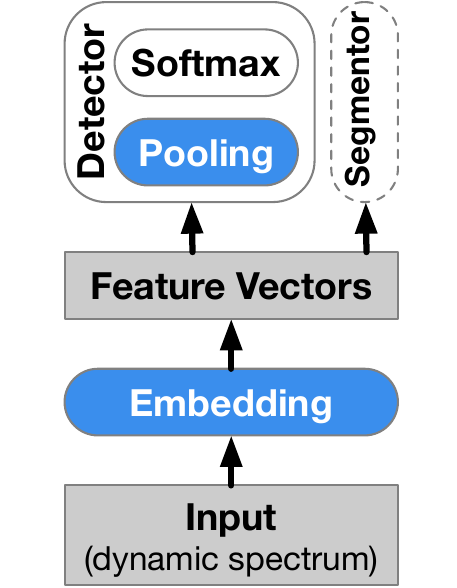}
    \caption{Framework of FRB detection models.}
    \label{fig:model-framework}
\end{figure}

All AI-based FRB detection models in this discussion can be described by a unified framework, as illustrated in Fig.~\ref{fig:model-framework}. The pipeline begins with an embedding module, which processes the input dynamic spectrum to generate a set of feature vectors. These features are then passed to a `Detector' module that determines whether the input contains an FRB.
The `Detector' incorporates a `Pooling' component to reduce the dimensionality of the feature vectors, facilitating efficient decision-making.
Optionally, if an FRB is detected, a segmentation module isolates the FRB signal from background noise.

In this work, we evaluate combinations of embedding architectures and feature aggregation (pooling) methods to identify optimal model structures for FRB detection.
We employ five embedding models, including VGG-19~\citep{Simonyan:2014cmh}, U-Net~\citep{ronnebergerUnetConvolutionalNetworks2015}, ResNet-18 \citep{heDeepResidualLearning2015}, and Swin UNETR \citep{hatamizadehSwinUNETRSwin2021}, along with four pooling strategies: global max/mean pooling, a four-layer Transformer encoder, and a TopK+Transformer module.
The Transformer encoder pooling module takes the feature map as a flattened sequence of feature vectors and aggregates the information into the special \texttt{[CLS]} token, and optionally into the \texttt{[DM]} token if the \texttt{dm} task is enabled.
The feature vectors corresponding to these special tokens are defined as the pooling output.
The concept of incorporating special tokens for pooling is inspired by the BERT model~\citep{devlinBERTPretrainingDeep2018}.
The TopK+Transformer pooling strategy follows the same procedure, except it operates only on the most salient feature vectors selected by the TopK module, rather than the entire feature map.

To save computation and training time, all models in this experiment are trained using 1 million simulated examples.
The model variants we constructed and their benchmark results on the FAST-FREX dataset are summarized in Table~\ref{tab:ablation-results}.

\begingroup
\setlength{\tabcolsep}{4pt}
\begin{table}[!tbhp]
\centering
\caption{Performance of different models on the FAST-FREX dataset.
\texttt{cls} is a mandatory task in our setting. The plus sign prefix of a task indicate one or more auxiliary tasks are enabled.
The \texttt{seg} task is also mandatory for architectures with inherent segmentation capabilities (U-Net, Swin UNETR).
$\texttt{psg} = \{\texttt{pcls}, \texttt{seg}\}, \texttt{all} = \{\texttt{dm}, \texttt{pcls}, \texttt{seg}\}. $
}
\begin{tabular}{llcrrr}
\toprule
Embedding  & Pooling & Task & \multicolumn{1}{c}{Rec.} & \multicolumn{1}{c}{Prec.} & \multicolumn{1}{c}{F1} \\
\midrule
\multirow[t]{3}{*}{VGG-19}      & Maxpool     & \texttt{cls}   &  7.5 & 100.0 & 14.0 \\
                                & Avgpool     & \texttt{cls}   & 17.0 & 98.1  & 29.0 \\
                                & Transformer & \texttt{cls}   & 25.0 & 94.9  & 39.6 \\
\multirow[t]{3}{*}{U-Net}       & Maxpool     & \texttt{+seg}  &  9.2 & 100.0 & 18.3 \\
                                & Avgpool     & \texttt{+seg}  & 18.3 & 98.2  & 30.1 \\
                                & Transformer & \texttt{+seg}  & 29.2 & 97.1  & 44.9 \\
\multirow[t]{3}{*}{ResNet-18}   & Maxpool     & \texttt{cls}   & 40.5 & 97.0  & 80.6 \\
                                & Avgpool     & \texttt{cls}   & 44.7 & 96.6  & 88.9 \\
                                & Transformer & \texttt{cls}   & 48.5 & 98.3  & 64.9 \\
\multirow[t]{4}{*}{Swin Trans.} & Avgpool     & \texttt{cls}   & 89.5 & 97.7  & 93.4 \\
                                & Transformer & \texttt{cls}   & 92.6 & 98.8  & 95.6 \\
                                & Transformer & \texttt{+pcls} & 93.5 & 99.0  & 96.2 \\
                                & TopK+Trans. & \texttt{+pcls} & 93.3 & 99.0  & 96.1 \\
\multirow[t]{5}{*}{Swin UNETR}  & Avgpool     & \texttt{+seg}  & 89.8 & 97.8  & 93.7 \\
                                & Transformer & \texttt{+seg}  & 92.8 & 98.9  & 95.8 \\
                                & Transformer & \texttt{+psg}  & 94.3 & 99.5  & 96.8 \\
                                & TopK+Trans. & \texttt{+psg}  & 94.8 & 99.0  & 96.9 \\
                                & TopK+Trans. & \texttt{+all}  & 94.7 & 99.3  & 96.9 \\
\bottomrule
\end{tabular}
\label{tab:ablation-results}
\end{table}
\endgroup

Our experiments reveal three critical findings about FRB detection architectures:

First, the choice of embedding network substantially impacts detection performance. Traditional CNNs like VGG-19 prove inadequate, while residual networks (ResNet-18) and Swin transformers achieve dramatically better results. The Swin transformers and Swin UNETR variants demonstrate particular effectiveness.

Second, feature aggregation strategy significantly affects model performance.
While conventional pooling methods (Maxpool, Avgpool) yield limited recall (7.5\%-89.8\%), transformer-based pooling improves detection rates by 3-20\% absolute.
The TopK+Transformer hybrid achieves the best or comparable performance in both efficiency and accuracy
by focusing computation on the most informative time-frequency regions.

Third, joint detection-segmentation training (\texttt{+pcls}) consistently outperforms pure classification. This is because the \texttt{pcls} task introduces additional supervision at deeper layers of the network, encouraging the learning of more discriminative and spatially aware feature representations.
The full pipeline (`\texttt{+all}') incorporating all auxiliary tasks achieves peak performance ($\text{F1} = 96.9\%$), though with diminishing returns over the segmentation-enhanced variant.

Based on our experimental findings, we select the Swin UNETR architecture with TopK+Transformer pooling as our final model, given its superior performance in FRB detection tasks.

\paragraph{FRB Simulator variants}
As described in Section~\ref{sec:frb-simulation}, our FRB simulator extends the \texttt{fitburst} package with two key components: a \textit{background sampler} and a \textit{blinking sampler}.
We systematically evaluate these components through controlled experiments measuring their impact on model performance.
The feature components are added to \texttt{fitburst} gradually to form different simulator variants.
We compare the performance of trained models based on these variants in Table~\ref{tab:simulator-ablation}.

\begin{table}[!tbhp]
\centering%
\caption{%
Impact of FRB simulator components on detection performance. %
The baseline \texttt{fitburst} implementation uses Gaussian noise, while extended versions incorporate: %
\texttt{+obs} (observation-based background sampler) and %
\texttt{+bs} (blinking sampler). %
Performance is measured by recall (Rec.), precision (Prec.), and F1 score on the FAST-FREX dataset (\%). %
}%

\begin{tabular}{lccc}%
\toprule%
Variants  & Rec. & Prec. & F1 \\%
\midrule%
\texttt{fitburst}  & 55.3 & 28.7 & 37.8 \\%
\texttt{  +bs}     & 61.1 & 24.8 & 35.3 \\%
\texttt{  +obs}    & 89.7 & 98.6 & 93.9 \\%
\texttt{  +obs+bs} & 94.7 & 99.3 & 96.9 \\%
\bottomrule%
\end{tabular}%
\label{tab:simulator-ablation}%
\end{table}%

Our experiments reveal critical insights about simulator design. Firstly, the observation-based background sampler (\texttt{+obs}) alone accounts for 95\% of the total F1 improvement ($37.8\% \rightarrow 93.9\%$ vs $37.8\% \rightarrow 96.9\%$), demonstrating that realistic background simulation is essential for both sensitivity (recall $+56.1\%$) and specificity (precision $+69.9\%$). Secondly, while the blinking sampler (\texttt{+bs}) alone degrades performance in F1 score ($37.8\%\rightarrow 35.3\%$), its combination with \texttt{+obs} yields additional gains ($93.9\% \rightarrow 96.9\%$).
It shows the limitations of modeling only scintillation without proper background, however,
scintillation modeling significantly improves sensitivity in FRB detection (recall $55.3\% \rightarrow 61.1\%$).
The current scintillation sampler relies on heuristic mathematical transformations rather than physics-based modeling, making it unclear whether this simplification limits further improvements in model performance.

These results validate our simulator design choices while identifying directions for future refinement.

\section{Applications}
\subsection{Blind Search on CRAFTS} \label{sec:blind-search-application}

\begin{figure*}[tbp!]
    \centering
    \begin{subfigure}[t]{\linewidth}
        \centering
        \includegraphics[width=0.95\linewidth]{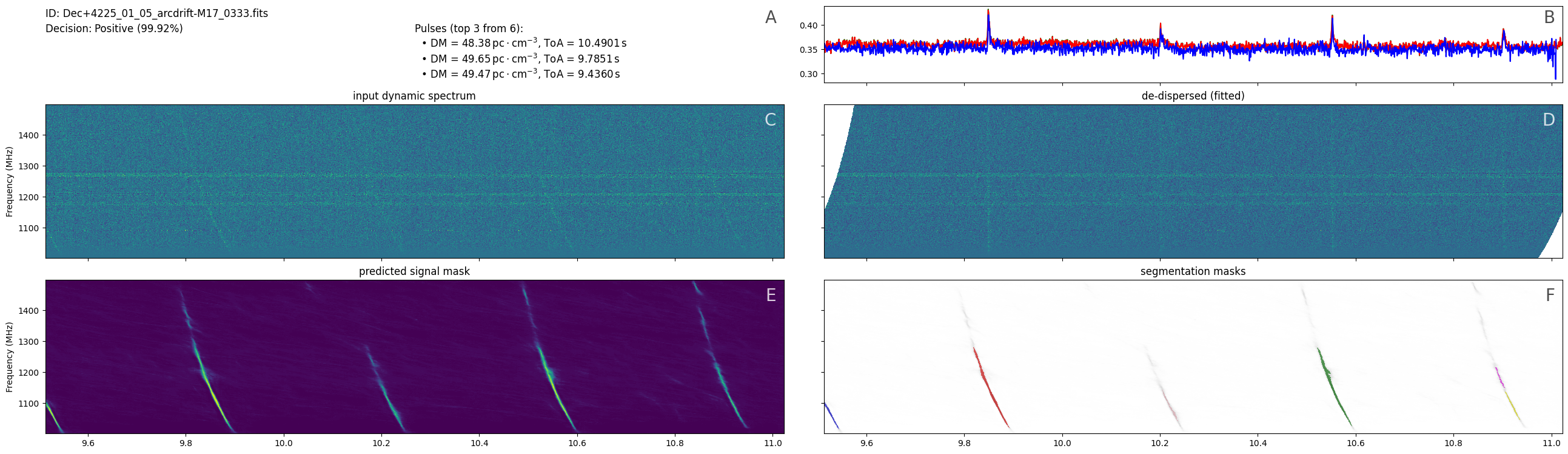}
        \caption{Detection report for a representative data segment from the first pulsar candidate detected by our model during the blind search on \texttt{CRAFTS ZD2024\_1\_1}.  Panels display: (A) detection summary, including input file name, classification result, and fitted DM and ToA parameters; (B) integrated pulse profiles; (C) input dynamic spectrum; (D) dedispersed dynamic spectrum using the predicted DM; (E) model-generated probability mask highlighting signal regions; and (F) instance-segmented signal masks.
        The report clearly reveals the pulsar pulses, though some single pulses may be erroneously segmented into multiple components.
        }
        \label{fig:1st-cand-model-predicted}
    \end{subfigure}
    \begin{subfigure}[b]{\linewidth}
        \centering
        \includegraphics[width=0.90\linewidth,trim={7ex 5ex 10ex 10ex},clip]{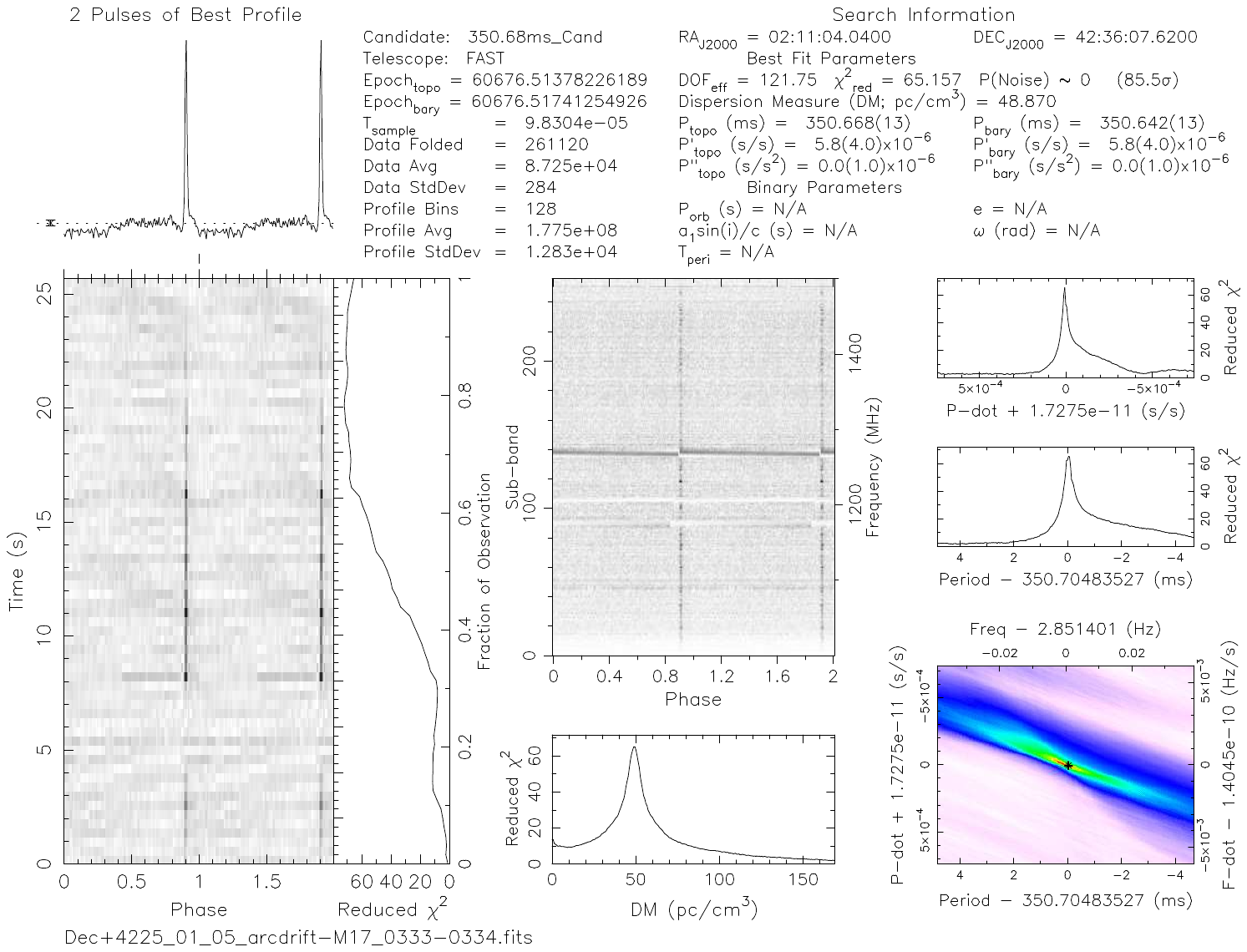}
        \caption{Diagnostic plot for the pulsar candidate shown in Fig.~\ref{fig:1st-cand-model-predicted}, generated using the \texttt{prepfold} utility from the \textsc{PRESTO} suite.
        The initial \texttt{prepfold} parameters, period ($P_0 = 350.86$\,ms) and dispersion measure ($\text{DM}_0 = 48.9\,\text{pc}/\text{cm}^3$), were derived from our model's predictions.
        The discrepancies between our model's predictions and the \texttt{prepfold} best-fit parameters are remarkably small: $\varepsilon_P = 0.192$\,ms and $\varepsilon_{\text{DM}} = 0.03\,\text{pc}/\text{cm}^3$.
        This candidate is identified as a single known pulsar, corresponding to \texttt{J0211+4233} from the CRAFTS survey \citep{li2018fast} and \texttt{J0211+4235} from the ATNF survey \citep{Manchester:2004bp}.
        }
        \label{fig:1st-cand-presto-profile}
    \end{subfigure}%
    \caption{Pulsar \texttt{J0211+4233} detected by our model on the \texttt{ZD2024\_1\_1/Dec+4225\_01\_05} subset from CRAFTS. }
    \label{fig:1st-cand}
\end{figure*}

\begin{figure*}[tbp!]
    \centering
    \begin{subfigure}[t]{\linewidth}
        \centering
        \includegraphics[width=0.95\linewidth]{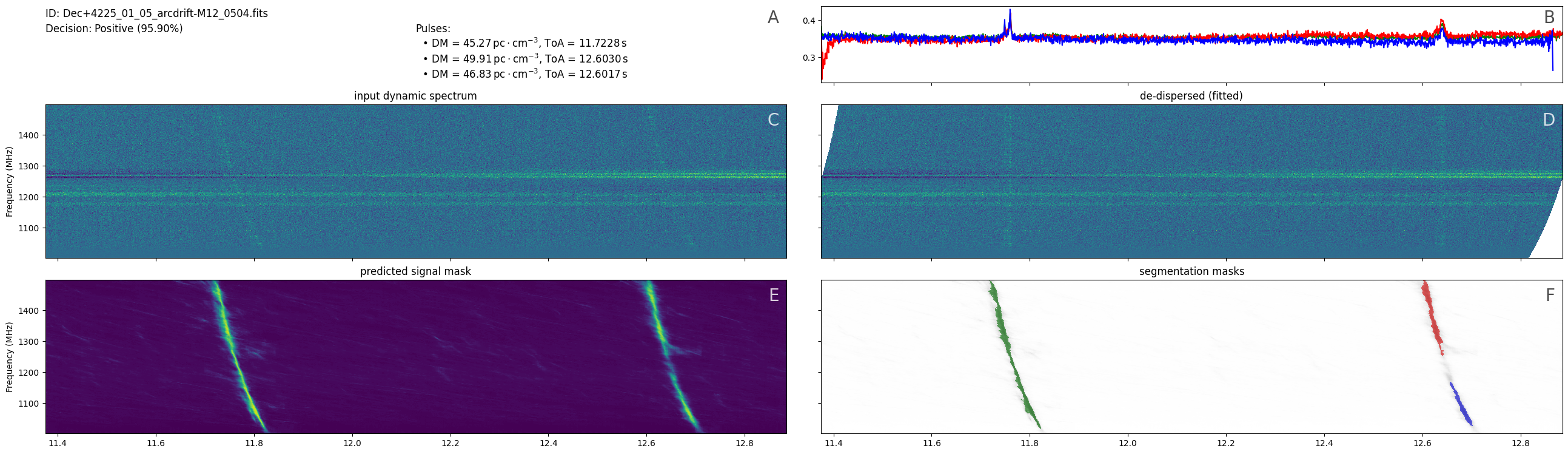}
        \caption{Detection report for a representative data segment from the second pulsar candidate detected by our model during the blind search on \texttt{CRAFTS ZD2024\_1\_1}.  Panels display: (A) detection summary, including input file name, classification result, and fitted DM and ToA parameters; (B) integrated pulse profiles; (C) input dynamic spectrum; (D) dedispersed dynamic spectrum using the predicted DM; (E) model-generated probability mask highlighting signal regions; and (F) instance-segmented signal masks.
        The report clearly reveals the pulsar pulses, though some single pulses may be erroneously segmented into multiple components.}
        \label{fig:2st-cand-model-predicted}
    \end{subfigure}
    \\
    \begin{subfigure}[b]{\linewidth}
        \centering
        \includegraphics[width=0.9\linewidth,trim={7ex 5ex 10ex 10ex},clip]{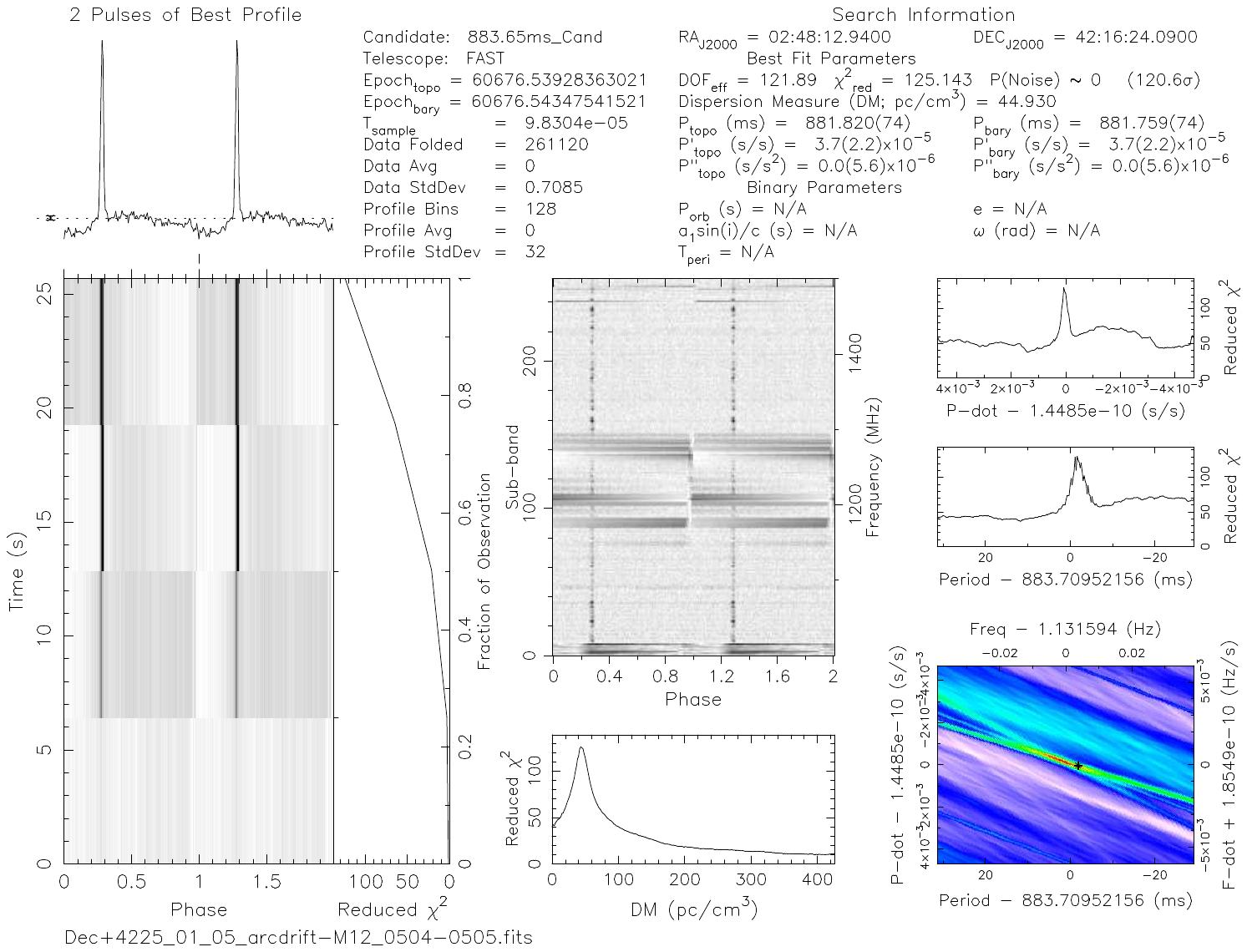}
        \caption{Diagnostic plot for the pulsar candidate shown in Fig.~\ref{fig:2st-cand-model-predicted}, generated using the \texttt{prepfold} utility from the \textsc{PRESTO} suite.
        The initial \texttt{prepfold} parameters, period ($P_0 = 883.65$\,ms) and dispersion measure ($\text{DM}_0 = 45.95\,\text{pc}/\text{cm}^3$), were derived from our model's predictions.
        The discrepancies between our model's predictions and the \texttt{prepfold} best-fit parameters are remarkably small: $\varepsilon_P = 1.83$\,ms and $\varepsilon_{\text{DM}} = 1.02\,\text{pc}/\text{cm}^3$.
        This candidate is identified as a single known pulsar, corresponding to \texttt{J0248+4220g} from the FAST-GPPS survey \citep{Wang:2024hjy} and \texttt{J0248+4220\_P} from the ATNF survey \citep{Manchester:2004bp}. }
        \label{fig:2st-cand-presto-profile}
    \end{subfigure}%
    \caption{Pulsar \texttt{J0248+4220\_P} detected by our model on the \texttt{ZD2024\_1\_1/Dec+4225\_01\_05} subset from CRAFTS. }
    \label{fig:2nd-cand}
\end{figure*}

We conducted an ongoing blind search for FRBs and radio pulsars using data from the \texttt{ZD2024\_1\_1} project of the CRAFTS survey \citep{li2018fast}. This project currently comprises 53 declination strips, each containing approximately 54~TB of observational data, amounting to a total of 2.8~PB. The results presented in this paper are based on the completed 40\% subset of the dataset.

Our model was first applied to process the FITS files according to the inference pipeline described in Section~\ref{sec:training-inference-method}.
All retrieved positive segments were subsequently subjected to manual verification to determine whether they represented true or false positives.
A FITS file is considered positive if at least one of its segments is classified as positive.

Table~\ref{tab:crafts-in-prod} presents the metrics for several randomly sampled declination strips and a specific declination strip that successfully identified positive candidates.
These results demonstrate that our model achieves a very low false positive rate (an average of 0.28\% and consistently below 0.5\% across all declination strips), significantly reducing the burden of manual verification.

In addition, we optimized the processing pipeline by overlapping I/O, pre-processing, and post-processing operations, achieving super-real-time data throughput.
This optimization resulted in an average processing rate of 0.63 seconds per second of observation.
Variations in processing speed across different declination strips are attributed to fluctuations in system load and network conditions, as the dataset is hosted on an Object Storage Service (OSS).
The combination of an exceptionally low false positive rate and high processing efficiency enables scalable blind searches on petabyte-scale datasets.

\begin{table}[!htb]%
\setlength{\tabcolsep}{3pt}
\centering
\caption{Blind search metrics of randomly sampled Dec strips and a positive candidate-recognizing Dec strip from the \texttt{CRAFTS ZD2024\_1\_1} project.
``TP'' = true positives, ``FP'' = false positives, ``FP'' = FP rate, and the time column is the average seconds to process one second of observations.}%
\label{tab:crafts-in-prod}%
\begin{tabular}{cccccc}%
\toprule%
Dec strips        &\#Files& TP& FP  & FPR ($\downarrow$) & time ($\downarrow$)\\%
\midrule%
Dec+4247\_01\_05 & 27436 & 0 & 85  & 0.31\% & 0.57\\%
Dec+5631\_06\_03 & 27436 & 0 & 41  & 0.15\% & 0.44\\%
Dec+5609\_06\_03 & 16815 & 0 & 53  & 0.32\% & 0.58\\%
Dec+2737\_07\_04 & 23009 & 0 & 113 & 0.49\% & 0.67\\%
Dec+0010\_09\_05 & 29659 & 0 & 64  & 0.22\% & 0.75\\%
Dec+4225\_01\_05 & 27436 & 9 & 76  & 0.28\% & 0.77\\%
\multicolumn{1}{r}{Total} & 124355& 9 & 356 & 0.28\% & 0.63\\%
\bottomrule%
\end{tabular}%
\end{table}%

Our model detected candidate signals in nine files from the \texttt{Dec+4225\_01\_05} strip.
Manual inspection of these files revealed that the detected pulses could be grouped into two adjacent sets of files.
Within each group, the pulses exhibited similar positive DM values and consistent ToA intervals between adjacent pulses.
In contrast, pulses from different groups showed clear differences in both right ascension (RA) and ToA separation.
These characteristics strongly suggest that the two groups correspond to signals from two distinct pulsar candidates. Figures~\ref{fig:1st-cand-model-predicted} and~\ref{fig:2st-cand-model-predicted} present the model's detection reports for representative segments corresponding to these candidates.
Both figures distinctly display periodic pulse signals characterized by positive dispersion measures.

To further validate our identified candidates, we conducted a follow-up analysis using PRESTO.
For each candidate, we first applied the \texttt{rfifind} command to generate a radio frequency interference (RFI) mask.
DM and ToA for individual pulses were then estimated by fitting Eq.~\eqref{eq:time-freq-dm}.
The overall DM of a pulsar candidate was determined by averaging the DMs of all its detected pulses.
Similarly, the candidate's period was estimated from the mean ToA intervals between adjacent pulses.
During this period estimation step, duplicate pulse detections were carefully removed. For instance, as shown in Fig.~\ref{fig:2st-cand-model-predicted} (panel F), a single pulse was erroneously segmented into two distinct predictions by our model due to strong RFI. The time separation between these spurious detections was not representative of the true period and was therefore must be excluded.
Finally, we performed a folding analysis using the \texttt{prepfold} command, incorporating the generated RFI mask, along with the estimated DM and period, as input parameters.


The results of the folding analysis for the two candidates are shown in Fig.~\ref{fig:1st-cand-presto-profile} and Fig.~\ref{fig:2st-cand-presto-profile}, respectively. In both figures, the top-left panel presents a highly significant and well-defined average pulse profile, indicating the presence of a strong periodic signal. The dynamic spectra (middle panels) demonstrate phase coherence of the pulses across both time and frequency, with no evident dispersion smearing, thereby confirming the correctness of the estimated DM.
Two prominent vertical bright bands in the dynamic spectra further validate the optimal folding period. Since the phase is displayed over two full cycles (from 0 to 2), the recurring pulse produces two clearly separated detections.
Additionally, the sharp positive peak in the DM trial plot (bottom-middle) provides further support for the derived DM value.
Complementary diagnostic plots on the right show well-constrained period and its derivatives, characterized by sharp peaks.
Together, these plots provide compelling evidence for a genuine astrophysical origin of the detected signals.

We extracted the right ascension, declination, best-fit period, and dispersion measure (DM) from the diagnostic plots Fig.~\ref{fig:1st-cand-presto-profile} and Fig.~\ref{fig:2st-cand-presto-profile}, and queried the pulsar database---\emph{Pulsar Survey Scraper} \citep{2022ascl.soft10001K}---to determine whether the candidates represent new discoveries.
The first candidate, shown in Fig.~\ref{fig:1st-cand}, matches the known pulsar \texttt{J0211+4233} from the CRAFTS survey \citep{li2018fast} and \texttt{J0211+4235} from the ATNF catalog \citep{Manchester:2004bp}.
The second candidate, shown in Fig.~\ref{fig:2nd-cand}, corresponds to \texttt{J0248+4220g} from the FAST-GPPS survey \citep{Wang:2024hjy} and \texttt{J0248+4220\_P} from the ATNF catalog.
These associations are supported by the close agreement in sky positions, pulse periods, and DMs.

Our blind search on the \texttt{CRAFTS ZD2024\_1\_1} dataset demonstrates that the proposed model enables efficient FRB and pulsar detection at the petabyte scale.
It achieves high robustness with minimal false positives, substantially reducing the need for manual verification.
This effectively overcomes the limitations of traditional methods, which rely heavily on labor-intensive manual verification.
Our model's extracted parameters are sufficiently precise for use as arguments in traditional analysis tools, such as \texttt{prepfold}, thus optimizing subsequent data processing.

\subsection{Integration with \texttt{fitburst} for Burst Fitting}

The previous application not only demonstrates our model's excellent performance in searching for FRBs or pulsars in a production environment,
but also showcases its effective integration with traditional astronomical tools.
Specifically, the DM and ToA values extracted from the model's output provide sufficiently precise input parameters for the \texttt{prepfold} command from PRESTO.

In this subsection, we further introduce an FRB fitting pipeline integrated with our model, serving to further showcase the effectiveness of our model's collaboration with traditional tools.
We integrate our model with the \texttt{fitburst} FRB fitting tool by using its predictions to initialize DM, ToA, and reference frequency (RF) parameters.
There are two model-based initialization methods to obtain the these parameters:
(i) \textit{Point-based}: The ToA and RF are estimated by extracting the center point coordinates from the segmentation mask, while the DM value is directly taken from the network’s \texttt{dm} output.
(ii) \textit{Fit-based}: The RF is derived from the center frequency of the segmentation mask, and both ToA and DM are estimated through a curve-fitting procedure applied to the extracted time-frequency coordinates, the same as the previous experiment.
For comparison, we introduce initializations based on the FAST-FREX dataset.
For these initializations, the RF is fixed to be 1500 MHz, and the DM and ToA parameters are taken from this dataset directly.

\begin{table}[!htbp]
    \centering
    \caption{Success rate of \texttt{fitburst} with different initialization strategies. ``Exc.'' refers to execution errors.}
    \label{tab:RoS-fitburst}
    \begin{tabular}{lccccc}
        \toprule
        Initialization & Success & Fail & Exc. & Total & RoS \\
        \midrule
        FAST-FREX   & 394     & 206  & 0    & 600   & 65.7\% \\
        fit-based   & 966     & 25   & 17   & 1008  & 95.8\% \\
        point-based & 964     & 27   & 17   & 1008  & 95.6\% \\
        \bottomrule
    \end{tabular}
\end{table}

Table~\ref{tab:RoS-fitburst} shows that model-based initializations boost the success rate of \texttt{fitburst} fitting from 65.7\% to over 95\%. Both fit-based and point-based parameter estimates yield similarly high RoS, illustrating the practical value of our model for enhancing usability in traditional analysis pipelines.

\begin{figure}[!htbp]
    \centering
    \includegraphics[width=0.99\linewidth]{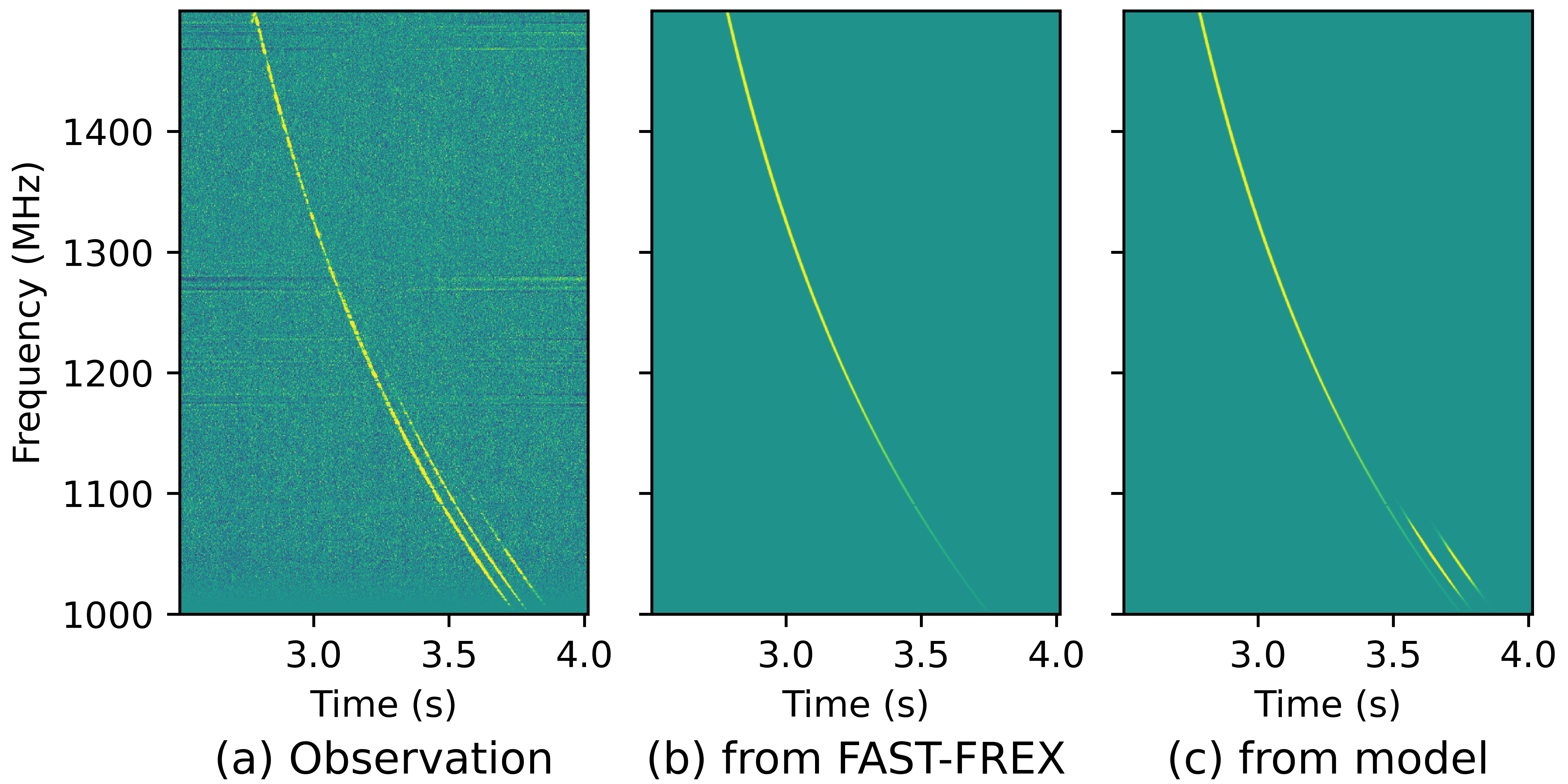}
    \caption{Example output of \texttt{fitburst} with different initializations.
    The model-initialized fitting successfully captures all three visible pulses, whereas the fitting initialized with FAST-FREX parameters recovers only a single pulse.}
    \label{fig:running-demo-fitburst}
\end{figure}

Fig.~\ref{fig:running-demo-fitburst} illustrates a representative example of the \texttt{fitburst} fitting results using different initialization strategies.
The fitting initialized with model-predicted parameters successfully captures all three pulse components present in the burst, whereas the baseline initialization based on FAST-FREX parameters identifies only the most prominent pulse, missing the others.
This highlights the effectiveness of our model's outputs in preserving fine-grained morphological structure of a burst,
which are critical for resolving complex burst structures.
Such improved tool may enhance not only fitting robustness but also the completeness of burst characterization, thereby facilitating more accurate physical interpretation and statistical analyses of FRB populations.


This application demonstrates our model's effective collaboration with traditional tools, enhancing the automation of FRB analysis. Parameters extracted from the model's output not only provide powerful priors or initialization for conventional tools but also enable more refined analysis.

\section{Limitations and Future Work}\label{sec:future_work}

Despite the promising results, several limitations in the evaluation and simulation constrain the full demonstration of our model's capabilities. Furthermore, architectural constraints regarding input resolution and the single-source assumption restrict the model's direct applicability across diverse observation settings.

\subsection{Limitations of the Evaluation Dataset}

The FAST-FREX evaluation dataset is limited in scale and diversity, containing only 600 positive examples from three FRB sources.
Critically, these sources are all repeaters, which represent only a minority of the total FRB population.
The absence of one-off bursts in our real-world testing set precludes a comparative performance analysis between these two distinct classes.
Furthermore, the lack of detailed annotations, such as signal masks or pulse substructures, prevents a fine-grained assessment of segmentation and parameter estimation.
While the model achieves 100\% precision on this dataset, this likely reflects the simplicity of the current evaluation setting rather than an absolute performance ceiling.

\textbf{Future Work:} We plan to expand the evaluation dataset by incorporating non-repeating FRB sources and richer annotations (e.g., sub-burst structures).
Adding a more diverse set of negative examples will also be essential to cover broader real-world noise variations and provide a more rigorous benchmark for detection and characterization tasks.

\subsection{Limitations of the Simulation Pipeline} \label{sec:simulator-limitations}

The current simulation pipeline, encompassing both background sampling and FRB signal generation, involves several simplifications that may affect model generalization.

\subsubsection{Background Sampling Constraints}
The background sampler assumes homogeneous noise characteristics, overlooking the heterogeneity arising from different telescope configurations and RFI environments. Moreover, the uniform sampling strategy is suboptimal, as it often draws low-information, trivial noise samples while underrepresenting rare but critical background or RFI patterns. This mismatch between training and real-world inference conditions can limit the model's robustness against complex noise artifacts.

\subsubsection{Signal Morphology and Physics}
Our simulator employs simplified templates that lack the fidelity to model complex structural variations such as sub-microsecond microstructure, quasi-periodic sub-bursts \citep{Pastor-Marazuela:2022pnp,Nimmo:2020sva}, or the characteristic downward frequency drift observed in repeating FRBs \citep{Hessels:2018mvq,Pleunis:2021qow}. Additionally, the assumption of a smooth spectral energy distribution fails to capture fine-scale frequency modulations like scintillation stripes \citep{Ravi:2016kfj} or plasma lensing fringes \citep{Chen:2024lop}.
While this has minimal impact on recall---which primarily depends on the $\nu^{-2}$ dispersion trace---it could compromise the accuracy of signal semantic segmentation and subsequent physical parameter estimation.

\textbf{Future Work:} To improve the simulation pipeline, we plan to implement a \textit{noise-aware sampling strategy} that uses observational metadata or unsupervised clustering to ensure more diverse background coverage. Furthermore, we intend to incorporate more flexible, physics-informed models into the simulator to capture complex signal morphologies, thereby enhancing the model's ability to generalize to the most sophisticated real-world detections.

\subsection{Architectural Constraints and Generalizability}

Our multi-task framework's direct parameter prediction imposes strict requirements on input resolution and spectral band,
breaking frequency-axis translation invariance. Consequently, a signal's DM is tied to its specific frequency coordinates,
leading to estimation errors if the input configuration deviates from the training preset.
While this sensitivity necessitates retraining for different telescopes,
the process is streamlined by our instrument-agnostic design and reliance on synthetic, label-free data.
Additionally, the model assumes a single DM per observation window, which may compromise accuracy during rare multi-source spatial overlaps.

\textbf{Future Work:} To enhance flexibility, we aim to decouple the input matrix from physical units by incorporating sampling metadata
as auxiliary model inputs. Alternatively, we may adopt a hybrid pipeline that utilizes the model's predicted signal masks for
post-processing parameter estimation, ensuring compatibility across diverse observational setups without retraining.

\section{Conclusion}\label{sec:con}

In this study, we presented a transformer-based multi-task deep learning framework that revolutionizes the FRB search and analysis pipeline by integrating detection, pixel-level segmentation, and parameter estimation into a single, end-to-end process. By training exclusively on synthetic data through a rule-based automated annotation pipeline, we have demonstrated that high-fidelity simulations can effectively bridge the domain gap to real-world observations.
This approach provides a robust solution to the fundamental lack of pixel-level annotations in radio astronomy, where the transient nature and complex morphologies of FRB signals make precise manual labeling practically unfeasible.

Our experimental results underscore the practical viability of this approach. The model achieved state-of-the-art performance on the FAST-FREX dataset with an F1 score of \textbf{97.8\%}, outperforming traditional heuristic-based tools as well as other deep learning models in both accuracy and inference speed. The successful deployment of the model in a petabyte-scale blind search—identifying pulsar candidates with a remarkably low false-positive rate of \textbf{0.28\%}—proves its reliability and scalability for next-generation, high-volume surveys. Furthermore, the provision of interpretable pixel-level masks and their seamless integration as initializations for tools like \texttt{fitburst} and \texttt{prepfold} offers a powerful bridge between deep learning precision and established astrophysical workflows.

While the current framework demonstrates exceptional robustness, its reliance on specific input resolutions and simplified signal templates suggests clear paths for future enhancement.
As discussed in the limitations, we aim to incorporate more complex morphological structured signals into our simulation pipeline to further refine semantic segmentation.
Additionally, developing a noise-aware sampling strategy and decoupling physical units from the input matrix will enhance the model’s adaptability across diverse telescope configurations without the need for extensive retraining. Ultimately, this work provides a scalable, efficient, and interpretable foundation for automated transient discovery in the era of big-data radio astronomy.

\begin{acknowledgments}

This work was supported by the National Key R\&D Program of China (2022YFB4501405),
the National Natural Science Foundation of China (NSFC, grant No.~12588202 and No.~12203045), and the Leading Innovation and Entrepreneurship Team of Zhejiang Province, China (Grant No.~2023R01008).
\end{acknowledgments}

\software{
fitburst \citep{fonsecaModelingMorphologyFast2024},
astropy \citep{Astropy:2022ucr},
PyTorch \citep{Paszke:2019xhz},
PRESTO \citep{ransomNewSearchTechniques2001},
riptide \citep{morelloOptimalPeriodicitySearching2020}
}
\bibliographystyle{aasjournal}
\bibliography{ref.bib}


\end{document}